\documentclass[11pt,a4paper,epsf,epsfig,psfrag]{article}
\usepackage{jheppub}
\usepackage{amsmath,graphicx,amsfonts, mathrsfs,amssymb}
\usepackage{amsmath,epsfig}
\usepackage{amssymb,amsfonts}
\usepackage{color}
\usepackage{latexsym}
\usepackage{epsfig}
\usepackage{pdfsync}
\usepackage{marvosym}
\usepackage{subfloat}
\usepackage{color}
\newbox\pippobox

\title{Gluodynamics and deconfinement phase transition under rotation from holography}

\author[a,b,1]{Xun Chen,\note{first author,}}
\author[b,]{Lin Zhang,}
\author[c,2]{Danning Li,\note{co-corresponding author.}}
\author[a]{Defu Hou, }
\author[b,3]{and Mei Huang \note{corresponding author.}}

\affiliation[a]{Institute of Particle Physics and Key Laboratory of Quark and Lepton Physics (MOS), Central China Normal University, Wuhan 430079, P.R. China}
\affiliation[b]{School of Nuclear Science and Technology, University of Chinese Academy of Sciences,\\Beijing 100049, P.R. China}
\affiliation[c]{Department of Physics and Siyuan Laboratory, Jinan University, Guangzhou 510632, P.R. China}
\emailAdd{chenxunhep@qq.com}
\emailAdd{zhanglin@ucas.ac.cn}
\emailAdd{lidanning@jnu.edu.cn}
\emailAdd{houdf@mail.ccnu.edu.cn}
\emailAdd{huangmei@ucas.ac.cn}

\abstract{
 We investigate rotating effect on deconfinement phase transition in an Einstein-Maxwell-Dilaton(EMD) model in bottom-up holographic QCD approach. By constructing a rotating black hole, which is supposed to be dual to rotating strongly coupled nuclear matter, we investigate the thermodynamic quantities, including entropy density, pressure, energy density, trace anomaly, sound speed and specific heat for both pure gluon system and two-flavor system under rotation. It is shown that those thermodynamic quantities would be enhanced by large angular velocity. Also, we extract the information of phase transition from those thermodynamic quantities, as well as the order parameter of deconfinement phase transition, i.e. the loop operators. It is shown that, in the $T - \omega$ plane, for two-flavor case with small chemical potential,  the phase transition is always crossover. The transition temperature decreases slowly with angular velocity and chemical potential. For pure gluon system with zero chemical potential, the phase transition is always first order, while at finite chemical potential a critical end point(CEP) will present in the $T - \omega$ plane.  }

\begin{document}
\maketitle
\section{Introduction}\label{sec-int}
One of the main goals of relativistic heavy ion collisions(RHIC) is to probe the possible transition from  color confined and chiral symmetry broken vacuum to a deconfined and chiral symmetric phase, named quark gluon plasma(QGP). It is of essential importance to clearly understand phase transitions from theory side, in order to specify the possible signals of phase transition from tremendous amount of experimental data.

Currently, from lattice simulations\cite{Aoki:2006we,Ding:2015ona}, it is widely accepted that without density effects the phase transition is a smooth crossover at temperature around $T\simeq0.150-0.170\rm{GeV}$. In relatively low energy collisions of heavy nuclei, a significant fraction of baryon numbers might be left in the fire ball after collision. In such cases, the effect of baryon number density $n_B$ or chemical potential $\mu_B$ could not be neglected. A first order transition is likely to appear at large $\mu_B$. In between them, there is a big possibility to see a second order transition point, the critical end point(CEP), which is the main goal of the recent beam energy scan project(BES)\cite{Aggarwal:2010cw,Odyniec:2013aaa,Luo:2017faz} in RHIC.

Besides, magnetic field and vorticity could also have significant impact on the hot/dense nuclear matter, especially in non-central collisions. The charged nucleons pass through each other at almost speed of light and such an large effective electric current might stimulate a very strong magnetic field in the collision centre. The simulation shows that such a magnetic field could reach up to $0.1\sim0.5\rm{GeV}$\cite{Skokov:2009qp,Voronyuk:2011jd,Bzdak:2011yy,Deng:2012pc}, though it only lasts for time scale of strong interaction. As well, for the non-central heavy ion collisions,  the angular momentum $J$  is shown to be in the range of $10^3 \hbar-10^5 \hbar$~\cite{Jiang:2016woz,Deng:2016gyh}. Both magnetic field and vorticity would significantly change the property of hot nuclear matter. The magnetic effect has been well studied from different methods, while the vortical effect attracts growing interests recently.

Besides the interesting anomalous effects, like the chiral vortical effect~\cite{Kharzeev:2007tn,Son:2009tf,Kharzeev:2010gr,Rajagopal:2015roa} and chiral vortical wave~\cite{Jiang:2015cva}, vorticity would significantly change the phase transition\cite{Chen:2015hfc,Jiang:2016wvv,Ebihara:2016fwa,Chernodub:2016kxh,Chernodub:2017ref,Wang:2018sur}, either the location or the transition order. The location of CEP might shift towards low temperature and large $\mu_B$ direction at large vorticity\cite{Jiang:2016wvv,Wang:2018sur}. Different from magnetic field, the angular velocity $\omega$ couples with spin(or angular momentum) uniquely for different charged particles. Thus, both the directions of spin of quarks and anti-quarks tend to be parallel to the angular velocity. Correspondingly, scalar pairing would be suppressed in rotating nuclear matter\cite{Jiang:2016wvv}.  As shown in \cite{Wang:2018sur}, there might be another critical end point in the $T-\omega$ plane. These studies are mainly based on Nambu--Jona-Lasinio(NJL) model, and the order parameters considered are mainly the chiral condensate $\langle\bar{q}q\rangle$. Thus, only chiral phase transition is relevant in those studies. However, the possibility of separation between the deconfinement phase transition and the chiral transition has been discussed \cite{McLerran:2007qj,McLerran:2008ua,Li:2018ygx}. A natural question would be how deconfinement phase transition behaves under rotation. Such kind of discussion is still quite limited in the literature.

It is quite interesting to see that both chiral phase transition and deconfinement phase transition could be well studied from holographic method, motivated by the discovery of the Anti-de Sitter/conformal field theory(AdS/CFT) correspondence\cite{Maldacena:1997re,Gubser:1998bc,Witten:1998qj}.  Besides the top-down brane systems\cite{Babington:2003vm,Kruczenski:2003be,Kruczenski:2003uq,Sakai:2004cn,Sakai:2005yt,Huang:2007fv,Abt:2019tas,Nakas:2020hyo}, in bottom up approach, the Einstein-Maxwell-dilaton(EMD) system\cite{Gubser:2008ny,Gubser:2008yx,DeWolfe:2010he,Gursoy:2007cb,Gursoy:2007er,Finazzo:2014cna,Zollner:2018uep,Ballon-Bayona:2020xls,Bohra:2020qom,Mamani:2020pks,He:2020fdi}provides a good starting point to investigate deconfinement phase transition, while the chiral phase transition\cite{Colangelo:2011sr,Dudal:2015wfn,Chelabi:2015cwn,Chelabi:2015gpc,Fang:2015ytf,Li:2016gfn,Li:2016smq,Bartz:2016ufc,Fang:2016nfj,Bartz:2017jku,Fang:2018vkp} and light meson spectra\cite{Gherghetta-Kapusta-Kelley,Gherghetta-Kapusta-Kelley-2,YLWu,YLWu-1,Li:2012ay,Li:2013oda,Colangelo:2008us,Ballon-Bayona:2020qpq,FolcoCapossoli:2019imm,Contreras:2018hbi} could be well described in the soft-wall model\cite{Karch:2006pv}. The order parameters of both  chiral transition and deconfining transition, chiral condensate $\langle\bar{q}q\rangle$ and expectation value of Polyakov loop $\langle L\rangle$ respectively, could be easily extracted from holographic dictionary. Coupling the two model actions together, it is possible to have the mutual interaction of these two kinds of dynamics, chiral and deconfinement phase transitions might be described simultaneously, for examples, see Refs.\cite{Chen:2019rez,Yang:2014bqa,Dudal:2017max,Cai:2012xh}. Therefore, it could be quite interesting to extend such a holographic study to consider the vortical effect on QCD phase transitions.

Then, we want to discuss the physical image of the deconfiment phase transition and heavy-quark probe. A standard confinement/deconfinement phases in the dual boundary QCD theory will lead to a first order Hawking/Page phase transition from thermal-AdS to black hole on the gravity side. As discussed in Ref.\cite{Yang:2015aia,Dudal:2017max}, a small/large black hole phase transition which corresponds to specious-confinement/deconfinement phases in the dual boundary theory. Notably, the boundary dual of the small black hole phase does not exactly correspond to confinement as it has a nonzero (albeit exponentially small) Polyakov loop expectation value while showing linear confinement for larger distances at low temperatures only. For this reason, we called this as specious-confined phase. Furthermore, the specious-confinement/deconfinement phase transition temperature decreases with the chemical potential and the corresponding first order small/large black hole phase transition line terminates at a second order critical point, as predicted by the well-known lattice study\cite{deForcrand:2002hgr}.

To introduce the rotating effect in holography, one has to solve the dual back gravity of rotating nuclear matter. In Refs.\cite{McInnes:2014haa,McInnes:2018mwj,McInnes:2018pmk,Arefeva:2020jvo,Erices:2017izj}, the authors take the Kerr-AdS black hole and study the rotating effect. Generally speaking, the Kerr-AdS black hole could be a good description at extremely high temperature, where the system tends to the conformal limit. At low temperature, certain non-conformal effect in the gravity system would be an important supplement. Since such kind of effect could be introduced by the dilaton field in the EMD system, we will try to construct the dual rotating hot and dense system in this model. In principle, one has to solve the equation of motion up to certain boundary condition, reflecting the rotating nuclear matter. In Refs.\cite{BravoGaete:2017dso,Sheykhi:2010pya,Awad:2002cz,Nadi:2019bqu}, the authors obtain the non-conformal rotating black hole solution by doing a coordinate transformation to the static black hole solution. We think this scenario might be a good first approximation, since it must be a solution to the Einstein equation, due to symmetry of the action. Thus, we will follow this scenario and try to obtain a rotating solution and study the vortical effect. To specify the exact boundary conditions and get a more realistic dual gravity background would be left for the future. Also, as a preliminary test of the rotating effect, we will focus on the deconfinement phase transition firstly.

The rest parts of this paper would be organized as follows. We will give a brief review of the EMD system and try to obtain the dual rotating black hole solution to hot and dense QCD matter in  Sec.\ref{sec-action}. Then, we investigate the rotation effect on deconfinement phase transition in pure gluon and two-flavor holographic QCD models in Sec.\ref{sec-phase}.  Finally, a brief summary would be given in Sec.\ref{sec-sum}.

\section{The Einstein-Maxwell-Dilaton system under rotation}
\label{sec-action}

As mentioned above, the EMD system could give good description of deconfinement phase transition. By self-consistently solving the equations of motion, one could obtain the dual gravity background, from which the thermal dynamical quantities could be extracted. It is shown that by carefully fitting the model parameter settings, one could get equation of states(EoS) comparable to lattice simulations. For the compactness of this work, we will briefly review the model first, and then discuss the extension to rotating case.

\subsection{The Einstein-Maxwell-Dilaton system}
Following Refs.\cite{DeWolfe:2010he}, the action of the EMD system takes the following form:

\begin{gather}
S =\frac{1}{16\pi G_5}\int d^5x\sqrt{-g}[R - \frac{h(\phi)}{4}F^2-\frac{1}{2}\partial_\mu\phi\partial^\mu\phi-V(\phi)]\label{S}.
\end{gather}
Here, $g$ is the determinant of metric $g_{\mu\nu}$. $G_5$ is the 5D Newton constant. $\phi$ is the dilaton field. $F_{\mu\nu}$ is the strength tensor of a $U(1)$ gauge field $A_\mu$, which corresponds to the baryon number current. $h(\phi)$ and $V(\phi)$ are kinetic gauge function and dilaton potential respectively, the form of which could be fixed by fitting the equation of state from lattice calculation. Generally,  if $F_{\mu\nu}\equiv0$, the above action would describe the case with zero chemical potential($\mu_B=0$). To consider finite baryon number density effect, the boundary value of time component of $A_\mu$ should be non-zero and $F_{\mu\nu}\neq 0$ correspondingly. From previous studies, the EMD system could describe gluo-dynamics well. The phase structure for light quark systems could be reproduced in such a framework\cite{Gubser:2008ny,Gubser:2008yx,DeWolfe:2010he,Gursoy:2007cb,Gursoy:2007er,Yang:2014bqa,Dudal:2017max,Cai:2012xh}. It could be extended to heavy flavor system, and give good description on QCD phase diagram in pure gluon system\cite{Cai:2012xh,Li:2011hp}, glueball spectra\cite{Li:2013oda}.

Generally, to extract the relevant physical quantities from holographic dictionary, one has to solve the dual background gravity first. Before going to rotating QCD matter, we will give a brief review on the holographic dual of QCD matter in thermodynamic equilibrium. Following previous studies, in this case, one could take the following metric ansatz
\begin{gather}
ds^2=\frac{L^2e^{2A_e(z)}}{z^2}[-G(z)dt^2+\frac{1}{G(z)}dz^2+d\vec{x}^2],
\end{gather}
with $L$ the AdS radius, $A_e$ a warp factor describing the deformation from AdS metric, and $G$ a function breaking the Lorentz symmetry at finite temperature. In principle, if  $h(\phi)$ and $V(\phi)$ are fixed, then one can solve $A_e$ and $G$ from the equation of motion(EoM). However, to guess the exact form of $V(\phi)$ might be complicated, while $A_e$ is phenomenologically relevant.

With the ansatz of metric above, we can get EoM from the action of the EMD system

\begin{equation}\label{field-phi}
\begin{aligned}
\phi^{\prime \prime} &+\phi^{\prime}\left(-\frac{3}{z}+\frac{G^{\prime}}{G}+3 A_e^{\prime}\right)-\frac{L^{2} e^{2 A_e}}{z^{2} G} \frac{\partial V}{\partial \phi} + \frac{z^{2} e^{-2 A_e} A_{t}^{\prime 2}}{2 L^{2} G} \frac{\partial h}{\partial \phi}=0,\\
\end{aligned}
\end{equation}

\begin{equation}
\begin{aligned}
A_{t}^{\prime \prime}+A_{t}^{\prime}\left(-\frac{1}{z}+\frac{h^{\prime}}{h}+A_e^{\prime}\right)=0,
\end{aligned}
\end{equation}

\begin{equation}G^{\prime \prime}+G^{\prime}\left(-\frac{3}{z}+3 A_e^{\prime}\right)-\frac{e^{-2 A_e} A_{t}^{\prime 2} z^{2} h}{L^{2}}=0,\end{equation}

\begin{equation}\label{dilaton-potential-Eq} \begin{aligned}
A_e^{\prime \prime} &+\frac{G^{\prime \prime}}{6 G}+A_e^{\prime}\left(-\frac{6}{z}+\frac{3 G^{\prime}}{2 G}\right)-\frac{1}{z}\left(-\frac{4}{z}+\frac{3 G^{\prime}}{2 G}\right)+3 A_e^{\prime 2} \\
&+\frac{L^{2} e^{2 A_e} V}{3 z^{2} G}=0,
\end{aligned}\end{equation}

\begin{equation}A_e^{\prime \prime}-A_e^{\prime}\left(-\frac{2}{z}+A_e^{\prime}\right)+\frac{\phi^{\prime 2}}{6}=0.\end{equation}

Generally, to solve the above equations, one has to take a certain form of the function $h(\phi)$ and the dilaton potential $V(\phi)$. As in Refs.\cite{Li:2011hp,Li:2012ay,Kajantie:2011nx}, it could be more convenient if one fixes $A_e$ and $h$ from QCD phenomena and solve $V(\phi)$ from the equation of motion. This scenario is also named potential reconstruction approach in the literature  \cite{Fang:2015ytf, Yang:2014bqa,Farakos:2009fx,Cai:2012eh}. In this way, $V(\phi)$ would be a temperature and/or chemical potential dependent quantity. We would emphasize that for a given functions of $A_e$ and $h$, the form of the dilaton potenial $V(\phi)$ can be fixed from the EOM  but with temperature dependent coefficients. In some sense, this is quite similar to the case in 4D finite temperature field theory,  the coefficients of interacting potential change with the medium thus are temperature dependent. Also, as shown in Refs.\cite{Fang:2015ytf,Kajantie:2011nx}, this scenario is a good approximation to the fixed potential approach. Therefore, we will follow this scenario in the current study. According to the study in \cite{Dudal:2017max}, the following deformed factor $A_e(z)$ and $h(\phi)$ could give good description of non-perturbative QCD phenomena
\begin{eqnarray}
A_e(z)&=&-\frac{3}{4} \ln \left(a z^{2}+1\right)+\frac{1}{2} \ln \left(b z^{3}+1\right)-\frac{3}{4} \ln \left(d z^{4}+1\right),\\
h(\phi(z))&=&e^{-c z^2-A_e(z)}.\label{eqf}
\end{eqnarray}
Therefore, we will follow this study and take the above form of $A_e, h$.  Here, $a,b,c,d$ are four model parameters, which can be fixed by lattice data of equation of state. The best fitting values of these parameters for two-flavor and pure gluon system are given in Table \ref{table:parameter}, in the upper line and lower line respectively. We will leave the details of fitting thermodynamic quantities to the next section.

\begin{table}[htbp]
	\centering
	\begin{tabular}{|c|c|c|c|c|c|c|}
		\hline  %
		\phantom&c&a&b&d&$G_5/L^3$&$T_c$ \\
		\hline
		$N_f = 2$&-0.227$GeV^2$&0.01$GeV^2$&0.045$GeV^3$&0.035$GeV^4$&1.1&211MeV \\
        \hline
		$N_f = 0$&1.16$GeV^2$&0.075$GeV^2$&0.12$GeV^3$&0.075$GeV^4$&1.2&265MeV \\
        \hline
	\end{tabular}
\caption{Fitted parameters and extracted corresponding critical temperatures from lattice results on thermodynamic properties of pure gluon system and two-flavor system, respectively.}  %
\label{table:parameter}
\end{table}

\begin{figure}
	\centering
	\includegraphics[width=15cm]{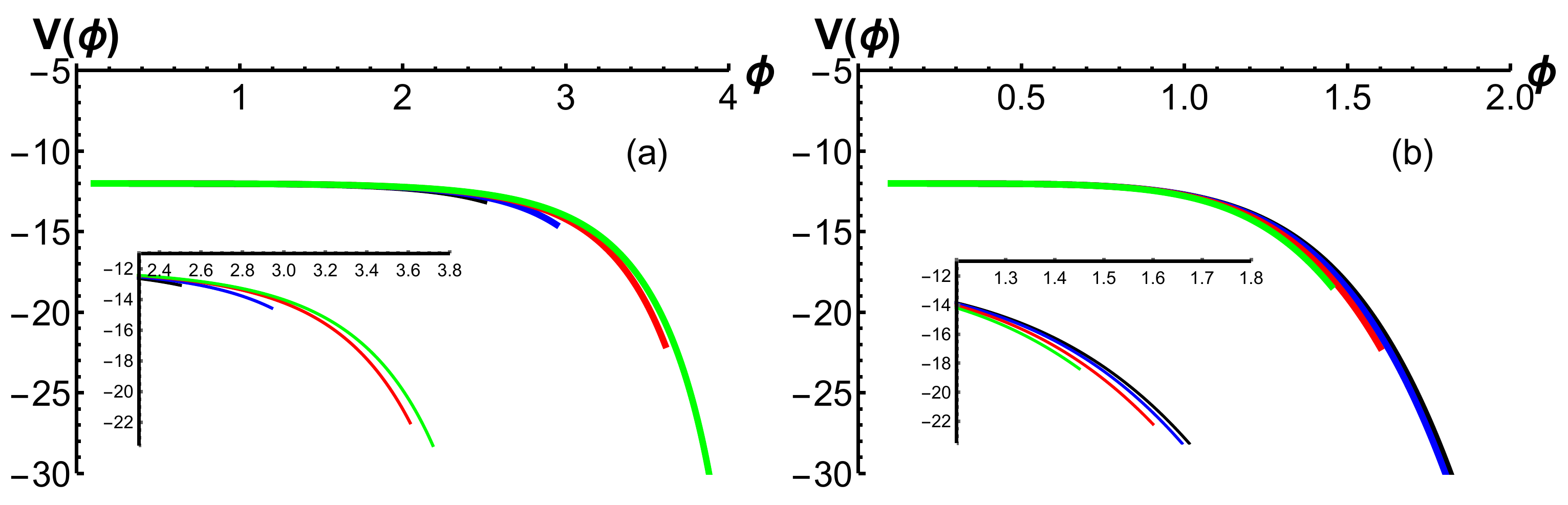}
	\caption{\label{Vphi} V($\phi$) as a function of $\phi$. (a) For different temperatures with $\mu = 0$. Black, blue, red and green lines are for $T=0.32{\rm GeV}$, $T=0.27{\rm GeV}$, $T=0.23{\rm GeV}$ and $T=0.21{\rm GeV}$, respectively. (b)For different $\mu$ with $T=0.21{\rm GeV}$. Black, blue, red and green lines are for $\mu=0$, $\mu=0.2GeV$,$\mu=0.4GeV$ and $\mu=0.6{\rm GeV}$,respectively. }
\end{figure}

\begin{figure}
	\centering
	\includegraphics[width=8cm]{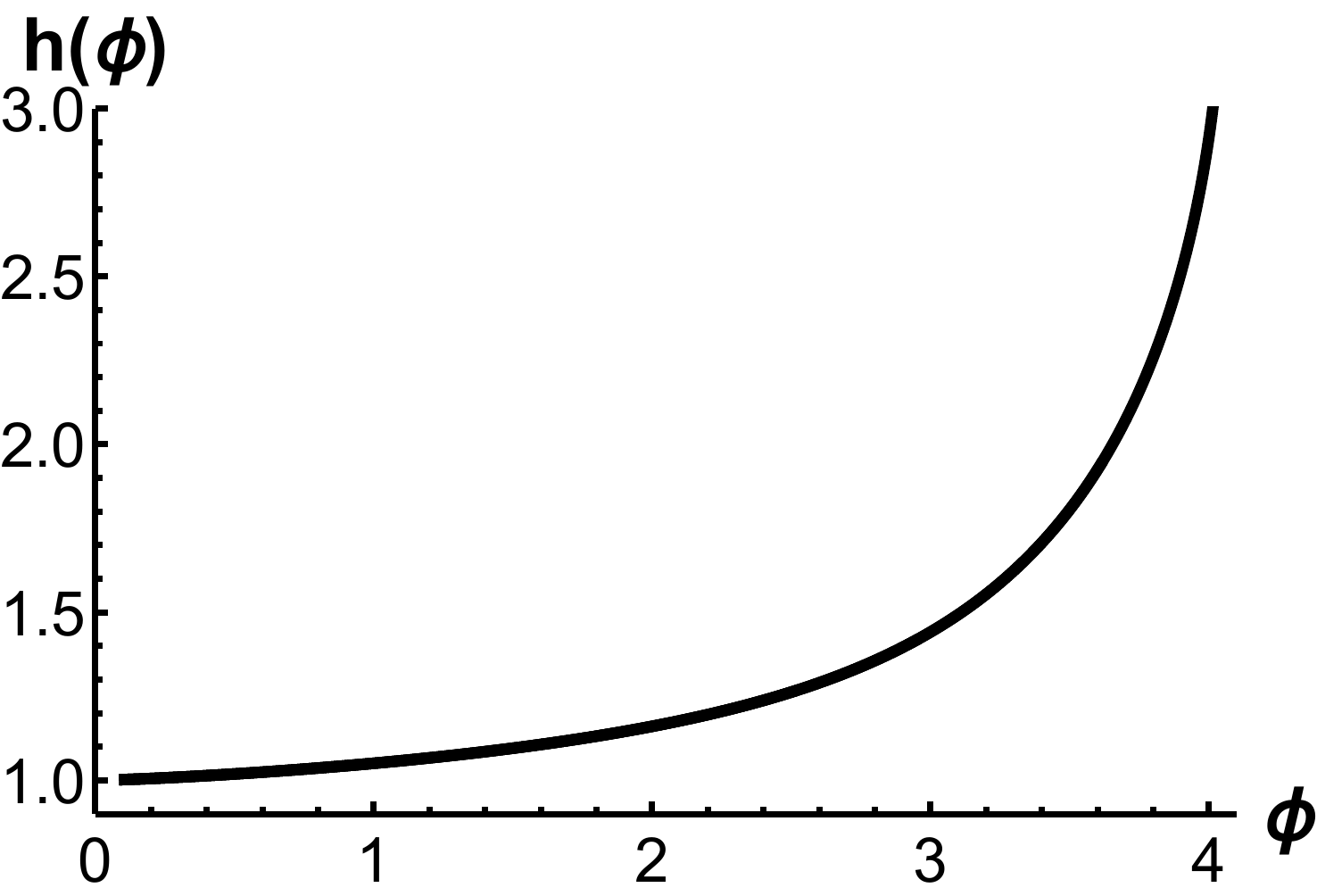}
	\caption{\label{hphi} h($\phi$) as a function of $\phi$. }
\end{figure}

Having the above form of $A_e$ and $h$, one can solve $G$ and $A_t$ from EoM. It has the following form\cite{Dudal:2017max}
\begin{eqnarray}
&&G(z)=1-\frac{1}{\int_{0}^{z_{h}} d x x^{3} e^{-3 A_e(x)}}\int_{0}^{z} d x x^{3} e^{-3 A_e(x)}\nonumber\\
&&+\frac{2 c \mu^{2}}{\left(1-e^{-c z_{h}^{2}}\right)^2\int_{0}^{z_{h}} d x x^{3} e^{-3 A_e(x)}} \left|\begin{array}{ll}
\int_{0}^{z_{h}} d x x^{3} e^{-3 A_e(x)} & \int_{0}^{z_{h}} d x x^{3} e^{-3 A_e(x)-c x^{2}} \\
\int_{z_{h}}^{z} d x x^{3} e^{-3 A_e(x)} & \int_{z_{h}}^{z} d x x^{3} e^{-3 A_e(x)-c x^{2}}
\end{array}\right|,\\
&&A_{t}(z)=\mu \frac{e^{-c z^{2}}-e^{-c z_{h}^{2}}}{1-e^{-c z_{h}^{2}}}.
\end{eqnarray}
Here, $z_h$ is the horizon of the black hole solution where $F(z_h)=0$, and $\mu$ is chemical potential. One can identify the temperature of 4D system as the Hawking temperature, which would be
\begin{eqnarray}
T=|\frac{G^\prime(z_h)}{4\pi}|.
\end{eqnarray}
 As mentioned above, the dilaton potential could be solved from Eq.\eqref{dilaton-potential-Eq}. We take different values of $z_h,\mu$ and show the difference of the dilaton potentials at different $T$ and $\mu$s in Fig.\ref{Vphi}(a) and (b) respectively. From the figure, we could see that the $T,\mu$ dependence of $V(\phi)$ is very weak and different lines almost merge together. Therefore, the current scenario could be a good approximation to the fixed potential approach.  We note that after $\phi(z)$ is solved, one can get the numerical form of $h(\phi)$ as a function of $\phi$. We plot $h(\phi)$ in Fig.\ref{hphi}. Different from $V(\phi)$, $h(\phi)$ is independent of $T,\mu$. The temperature dependence of $G(z)$ in Eqs.\eqref{field-phi},\eqref{dilaton-potential-Eq} are cancelled exactly by that of $V(\phi)$.

Also, after extracting the other thermodynamic quantities, like entropy, pressure, sound speed, one can obtain the information of phase transition of QCD. With the parameters in Table.\ref{table:parameter}, the phase transition is of first order type for pure gluon system and crossover type for two-flavor system. To calculate the phase transition temperature, we can use the order parameter(the expectation value of a single Polyakov loop) for the deconfinement transition in a pure $\text{SU}(N)$ theory and extract the transition temperature as $T_c=265\rm{MeV}$. When adding the dynamic quark, we still assume that Polyakov loop is an approximate order parameter. We also extract the pseudo-transition temperature of two-flavor system from this order parameter and it locates at $T_c=211\rm{MeV}$. The results of transition temperature are also summarized in Table.\ref{table:parameter}.

\subsection{Extend the EMD system to the rotating case}
To obtain the dual gravity background to rotating nuclear matter, a natural approach is to solve the equation of motion from the action in Eq.\eqref{S}, subject to the boundary condition for a rotating 4D system. However, an exact solution might not be easy to get. It is quite common that the distribution of the rotating system would depend on the distance to the rotating axis. Thus, one has to solve a partial derivative equations to get such a solution. Fortunately, if one only tries to get the qualitative behavior of the rotating matter at a fixed radius, the scenario used in Refs.\cite{Erices:2017izj,BravoGaete:2017dso,Sheykhi:2010pya,Awad:2002cz,Nadi:2019bqu} might provide a first approximation. According to Refs.\cite{Erices:2017izj,BravoGaete:2017dso,Sheykhi:2010pya,Awad:2002cz,Nadi:2019bqu}, one can get the rotating extension from the static configuration through a local Lorentz boost
\begin{eqnarray}
t \rightarrow \frac{1}{\sqrt{1-(\omega l)^2}}(t + \omega l^2 \theta), \phi \rightarrow \frac{1}{\sqrt{1-(\omega l)^2}}(\theta + \omega t).
\end{eqnarray}
Here, $l$ is the radius to the rotating axis. $\theta$ is the angular coordinate describing the rotation. $\omega$ is the angular velocity. Since we will focus on the qualitative results, we simply take $l=1\rm{GeV}^{-1}$. Then, the corresponding transformation of the metric would be
\begin{equation}
ds^2 = -N(z)dt^2 + \frac{H(z)dz^2}{G(z)} + R(z) (d\theta + P(z)dt)^2 + H(z)\sum\limits_{i=1}^2dx_i^2,
\end{equation}
with
\begin{align}
N(z) &= \frac{H(z)G(z)(1-\omega^2 l^2)}{1-G(z)\omega^2 l^2}, \\
H(z) &= \frac{L^2e^{2A_e(z)}}{z^2}, \\
R(z) &= H(z)\gamma^2 l^2 - H(z)G(z)\gamma^2\omega^2 l^4,  \\
P(z) &= \frac{\omega - G(z)\omega }{1 - G(z)\omega^2 l^2},\\
\gamma &= \frac{1}{\sqrt{1-\omega^2l^2}}.
\end{align}
Here, $G(z),A_e(z)$ are functions obtained in last section. For later convenience, we also extract the `metric' component $\hat{g}_{00}$(different from the $t-t$ component of metric $g_{00}$) and the $z-z$ component of the inverse of metric $g_{11}$ as
\begin{align}
\hat{g}_{00} &= -\frac{H(z)G(z)(1-\omega^2l^2)}{1-G(z)\omega^2l^2},\\
g^{11} &= \frac{G(z)}{H(z)}.
\end{align}
The Hawking temperature can be calculated from the surface gravity $\kappa$ as\cite{Zheng1983Hawking}
\begin{eqnarray}
T = \bigg|\frac{\kappa}{2\pi}\bigg| = \bigg|\frac{\lim\limits_{z\rightarrow z_h}-\frac{1}{2}\sqrt{\frac{g^{11}}{-\hat{g}_{00}}}\hat{g}_{00,1}}{2\pi}\bigg|.
\end{eqnarray}
The difference of the gauge potential is defined as chemical potential, namely, $\mu^{\prime}=\left.A_{t}\right|_{z=0}-\left.A_{t}\right|_{z=z_h}$\cite{Natsuume:2014sfa}. One thing we should notice is that the expression of the chemical potential is different from the static case because of the Lorentz transformation of time and angular component. Following Ref.\cite{Sheykhi:2010pya,Cvetic:1999ne,Caldarelli:1999xj}, the quark chemical potential under rotation can be defined as
\begin{eqnarray}
\mu=\left.A_{\mu} \chi^{\mu}\right|_{z = z_h}-\left.A_{\mu} \chi^{\mu}\right|_{z=0}.
\end{eqnarray}
Here, $\chi$ is the Killing vector taking the form
\begin{eqnarray}
\chi=\partial_{t}+ \Omega \partial_{\phi}.
\end{eqnarray}
The gauge field under the boost becomes
\begin{eqnarray}
A_{\mu}=\gamma A_t \delta^{t}_{\mu} + \omega \gamma A_t \delta^{\phi}_{\mu}.
\end{eqnarray}
The angular velocity of the black hole at the horizon is $\Omega = - P(z_h) = -\omega$\cite{Erices:2017izj,BravoGaete:2017dso}. After a short calculation, comparing with static case, the chemical potential would become
\begin{eqnarray}
\mu= \mu^{\prime}\sqrt{1-\omega^2}.
\label{muprime}
\end{eqnarray}

\begin{figure}
	\centering
	\includegraphics[width=15cm]{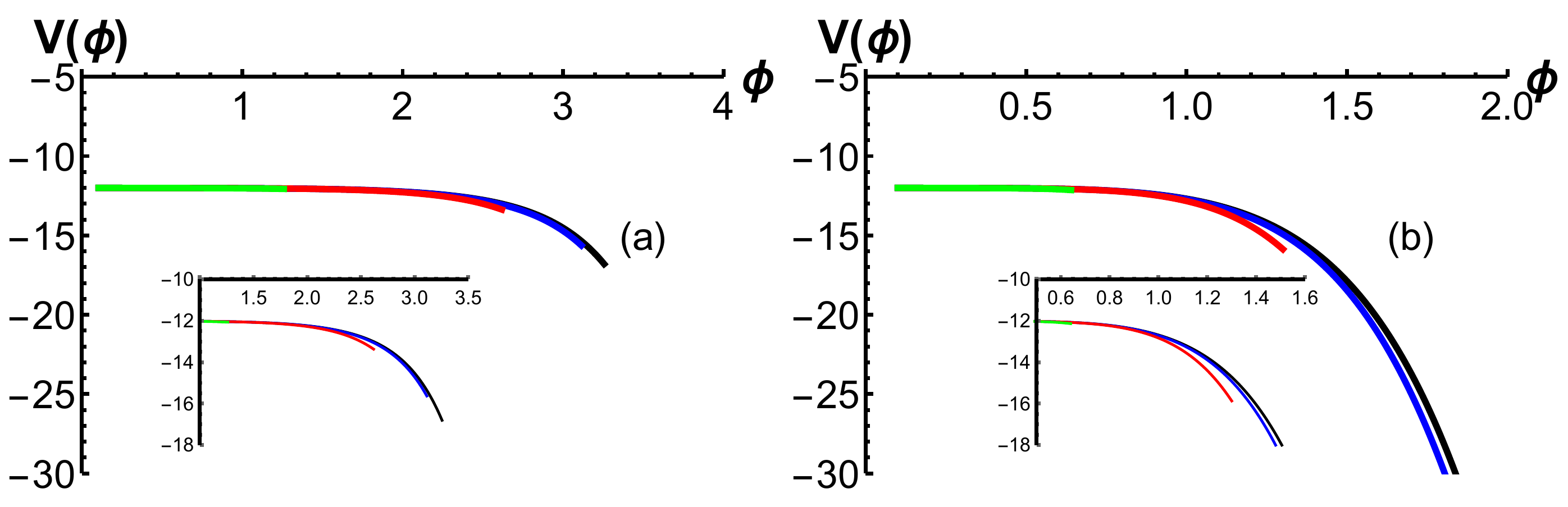}
	\caption{\label{Vphi2}  V($\phi$) as a function of $\phi$. (a) For different angular velocities with $\mu = 0$ and $T = 0.25{\rm GeV}$. Black, blue, red and green lines are for $\omega=0{\rm GeV}$, $\omega=0.3{\rm GeV}$, $\omega=0.6{\rm GeV}$ and $\omega=0.9{\rm GeV}$, respectively. (b)For different angular velocities with $\mu = 0.2{\rm GeV}$ and $T = 0.2{\rm GeV}$. Black, blue, red and green lines are for $\omega=0{\rm GeV}$, $\omega=0.3{\rm GeV}$, $\omega=0.6{\rm GeV}$ and $\omega=0.9{\rm GeV}$, respectively.} 
\end{figure}

 Note that, for numerical calculations, we always fix $\mu$ in our paper. At the end of this section, we also check the rotational effect on the dilaton potential in Fig.\ref{Vphi2}. It can be seen that angular velocity also has little effect on dilaton potential. 

\section{Gluodynamics under rotation}
\label{sec-phase}
With all the preparation in the above section, we will study the thermodynamics and phase transitions under rotation for pure gluon system and two-flavor system in this section. Firstly, we will try to investigate the behavior of thermodynamic quantities under rotation. Then we will examine the order parameter of deconfinement phase transition, i.e. the loop operators.

\subsection{Thermodynamic properties under rotation}
\label{sec:puregluon}

Generally, there are two ways to get the thermodynamic quantities. One can start from the free energy, i.e. the renormalized on-shell action, or from the Hawking-Beikenstein formula for entropy. In principle, the two scenario should be equivalent. There is certain uncertainty in adding (finite) counter terms in the former one, while in the latter one, there is uncertainty when integrating the entropy density. Therefore, it is possible to match the two scenario. For simplicity, we choose the latter one and start from the entropy density. From the Hawking-Beikenstein formula, the entropy density of a rotating black hole can be calculated as
\begin{eqnarray}\label{entropyformula}
s =\frac{A_{area}}{4G_5V_3}\bigg|_{z_h} = \frac{\sqrt{R(z_h)H(z_h)^2}}{4G_5}.
\end{eqnarray}
Here, we take $G_5/L^3 = 1.2$ by fitting lattice results of two-flavor system and $G_5/L^3 = 1.1$  for the pure gluon system\cite{Burger:2014xga,Boyd:1996bx}, as listed in Table.\ref{table:parameter}.
The free energy in the grand-canonical ensemble is given below \cite{Dehghani:2002rr}
\begin{eqnarray}
f = \epsilon - Ts - \mu \rho - \omega J.
\end{eqnarray}
Here $\epsilon$ is the energy density of the system and $J$ is the angular momentum per volume. From Refs.\cite{Jiang:2016woz,Deng:2016gyh}, we know that QCD matter will carry a local angular velocity in the range $0.01 \sim 0.1\rm{GeV}$. Thus, we will calculate EoS with a reasonable range $\omega = 0 \sim 0.2\rm{GeV}$. Variation in the free energy density of a system with a given volume can be written as
\begin{eqnarray}
df = - s dT - \rho d\mu -  J d\omega.
\end{eqnarray}
For fixed values of the chemical potential $\mu$ and angular velocity, the free energy can be computed by the following integration
\begin{eqnarray}
f = - \int s dT.
\end{eqnarray}
One can start from a given value of $\mu$ and $\omega$, and get free energy by integrating the entropy.

\begin{figure}
	\centering
	\includegraphics[width=15cm]{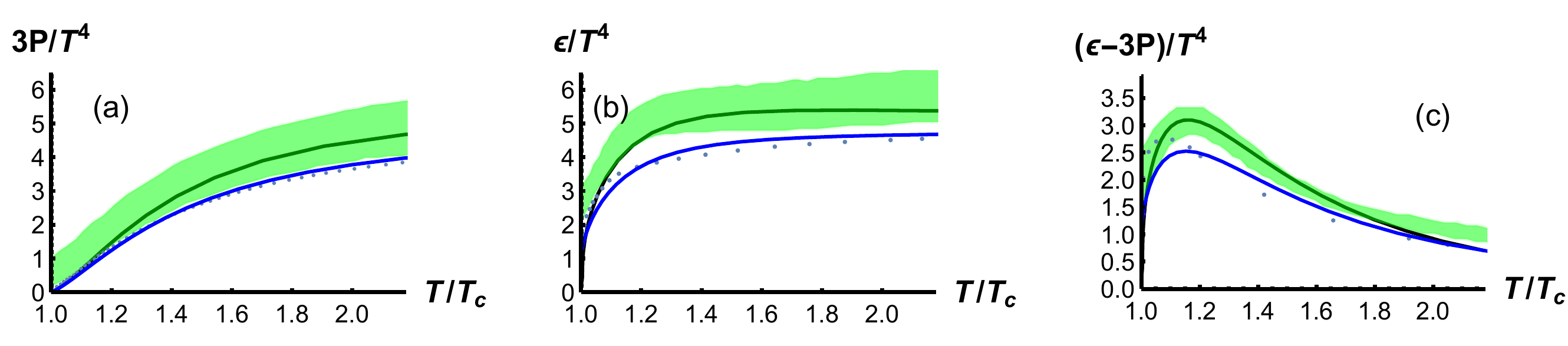}
	\caption{\label{eosall} Thermodynamic properties from lattice calculation and holographic QCD model with fitted parameters for pure gluon system and two-flavor system,respectively. (a) Pressure as a function of T. (b) Energy as a function of T. (c) trace anomaly as a function of T. Blue solid line is the results of holographic model and blue dot is lattice for pure gluon system both at vanishing chemical potential and angular velocity. Black solid line is the results of holographic model and green band is lattice for $N_f = 2$ both at vanishing chemical potential and angular velocity. }
\end{figure}

We will first consider the two-flavor case. Considering the vanishing chemical potential case, we set the pressure at $T_0 \approx 180 \rm{MeV}$ to be zero, close to lattice of two-flavor system in Ref.\cite{Burger:2014xga}. Requiring $f(T_0) = 0$ at $\mu,\omega = 0$, we are able to calculate the free energy density as

\begin{eqnarray}
f =  \int_{z_h}^{z_h(T_0)} s \frac{dT}{dz_h} dz_h.
\end{eqnarray}

Then one might extract the pressure as $P = -f$, the energy density as $\epsilon = f + s T + \mu \rho + \omega J$, and then all the other thermodynamic quantities. Firstly, we take $\omega=0,\mu=0$ and compare our results with lattice data in Fig.\ref{eosall}. In this figure, we show the pressure and energy density as a function of temperature. The solid black curve is our model result for two-flavor system. The green bands are lattice results from Ref.\cite{Burger:2014xga}. From the figure, we could see that at finite temperature, our model calculation agrees well with lattice results for two-flavor system.

\begin{figure}
	\centering
	\includegraphics[width=15cm]{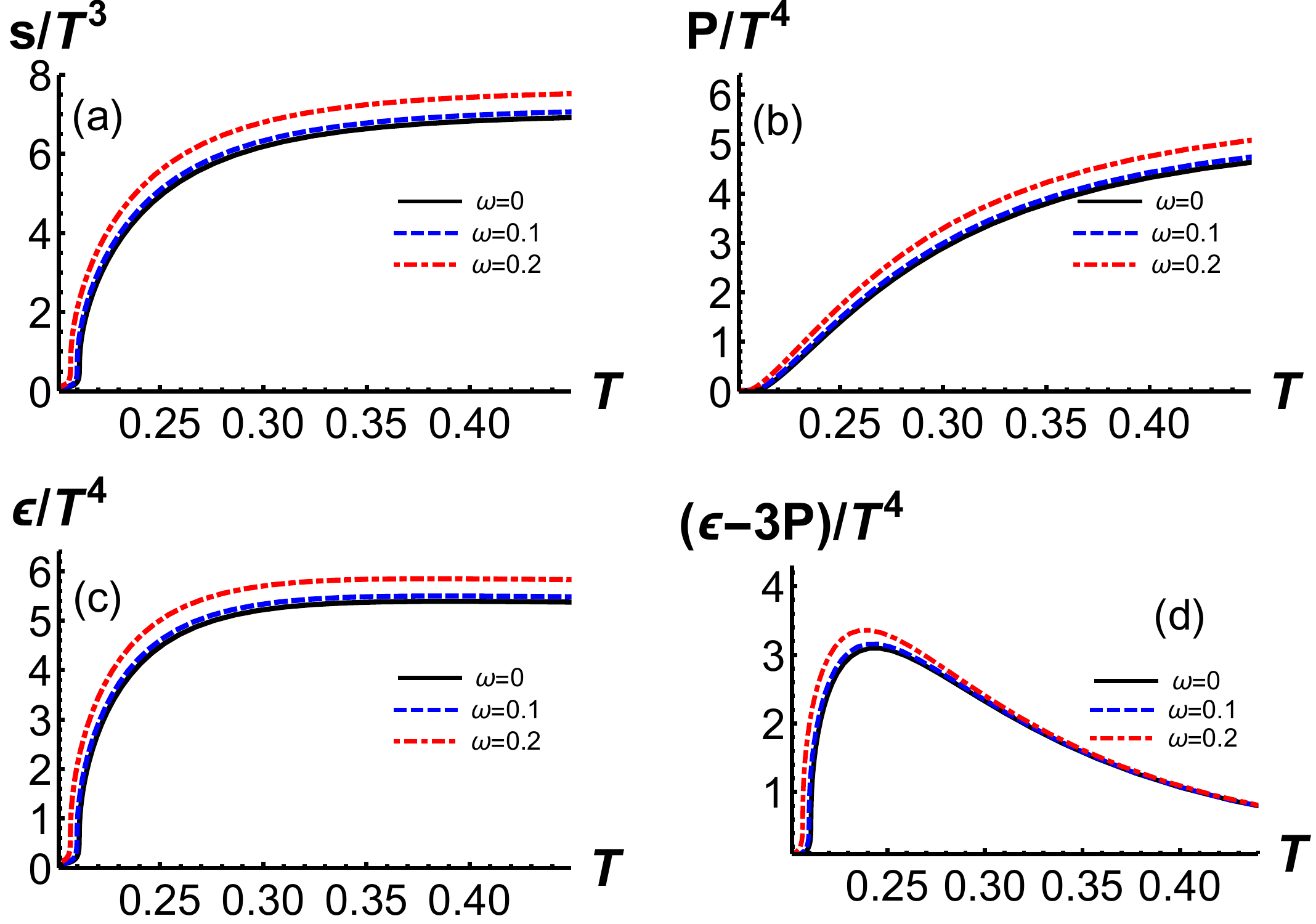}
	\caption{\label{lighteos} Thermodynamic properties of the two-flavor system for different angular velocities:(a) The entropy as a function of temperature for different angular velocities. (b) The pressure as a function of temperature for different angular velocities. (c)The energy as a function of temperature for different angular velocities. (d)The trace anomaly as a function of temperature for different angular velocities. Black solid line is $\omega = 0$, blue dashed line is $\omega = 0.1\rm{GeV}$ and red dot-dashed is $\omega = 0.2\rm{GeV}$ for vanishing chemical potential. The unit of $T,\omega$ is GeV.}
\end{figure}

\begin{figure}
	\centering
	\includegraphics[width=15cm]{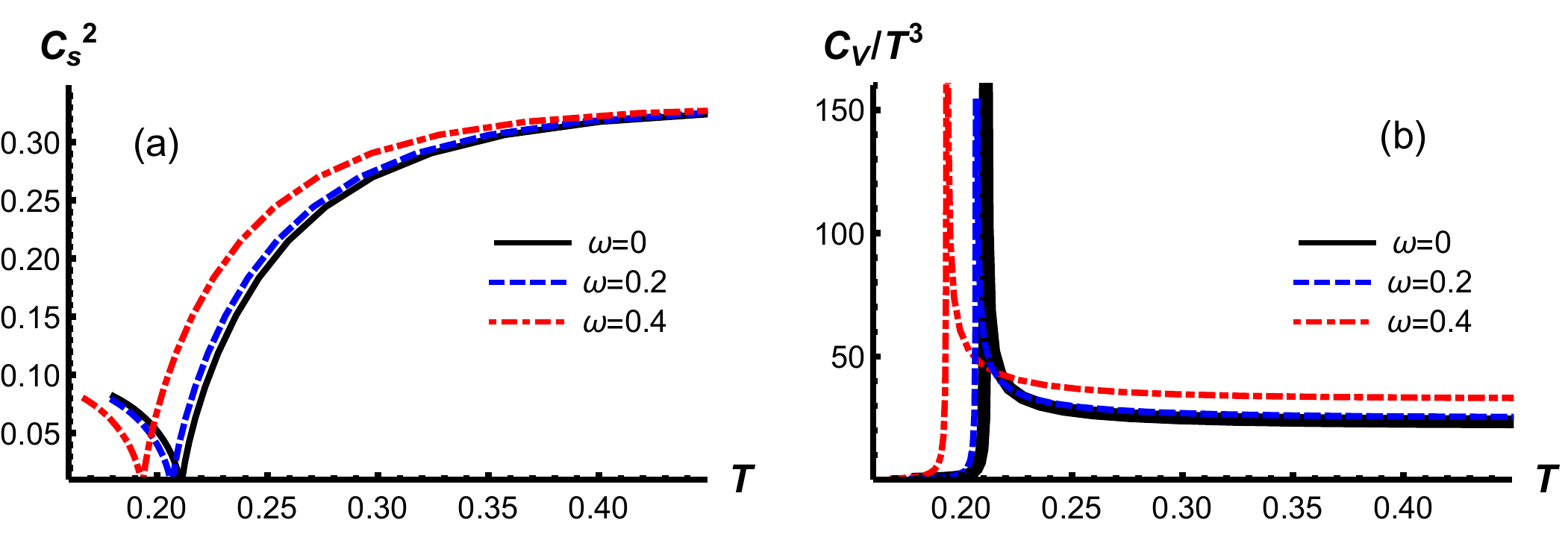}
	\caption{\label{lightcscv} For two-flavor system, The square of sound velocity (a) and the specific heat (b) as a function of temperature for different angular velocities with $\omega = 0$(black solid line), $\omega = 0.2\rm{GeV}$ (blue dashed line) and $\omega = 0.4\rm{GeV}$(red dot-dashed line). The unit of $T,\omega$ is GeV.}
\end{figure}

Then, we turn on the angular velocity and examine the effect of angular velocity on the EoS. We take $\omega=0, \omega=0.1\rm{GeV}, \omega=0.2\rm{GeV}$ and solve the scaled entropy, pressure, energy and trace anomaly as functions of $T$. The results are shown in Fig.\ref{lighteos}. It is shown that the scaled entropy, pressure and energy will increase with the increase of angular velocity and tend to constants at large $T$ limit. However, the trace anomaly for different values of $\omega$ will merge together at large $T$ limit. From Fig.\ref{lighteos}, those thermodynamic quantities would be enhanced by the angular velocity, almost $7\%$ from $\omega=0$ to $\omega=0.2\rm{GeV}$ in the large T limit. It is also observed that this behavior is quite different from the chemical potential effect\cite{Yang:2014bqa}.

Besides, the speed of sound $C_s^{2}$ and the specific heat $C_V$ also have information of phase transition. Both of these two quantities are related to derivative of entropy with respect to temperature
\begin{eqnarray}
C_s^{2} = \frac{\partial \ln T}{\partial \ln s}, C_V = T (\frac{\partial s}{\partial T}).
\end{eqnarray}

We show the behavior of the square of sound velocity and specific heat as a function of $T$ in Fig.\ref{lightcscv} for vanishing chemical potential. From this figure, the square of sound velocity will approach to the conformal limit $1/3$ for different angular velocities at large $T$ limit as expected. The temperature scaled specific heat $C_V/T^3$ will increase with the increase of angular velocity and change from $23.3$ for $\omega=0$ to $25.5,33.2$ for $\omega=0.1,0.2\rm{GeV}$ at large $T$ region.

As for pure gluon system, the calculation is slightly different from two-flavor system. Firstly, following Ref.\cite{Dudal:2017max}, we take $c=1.16 GeV^{2}$ by matching the holographic meson mass spectrum to that of lowest lying $J/\psi$ meson states. Again, we start from the free energy. Following Ref.\cite{He:2013qq}, we require $f\left(z_{h} \rightarrow \infty\right)=0$ and define

\begin{equation}
f=\int_{z_{h}}^{\infty} s \frac{d T}{d z_{h}} d z_{h}.
\end{equation}

\begin{figure}
	\centering
	\includegraphics[width=15cm]{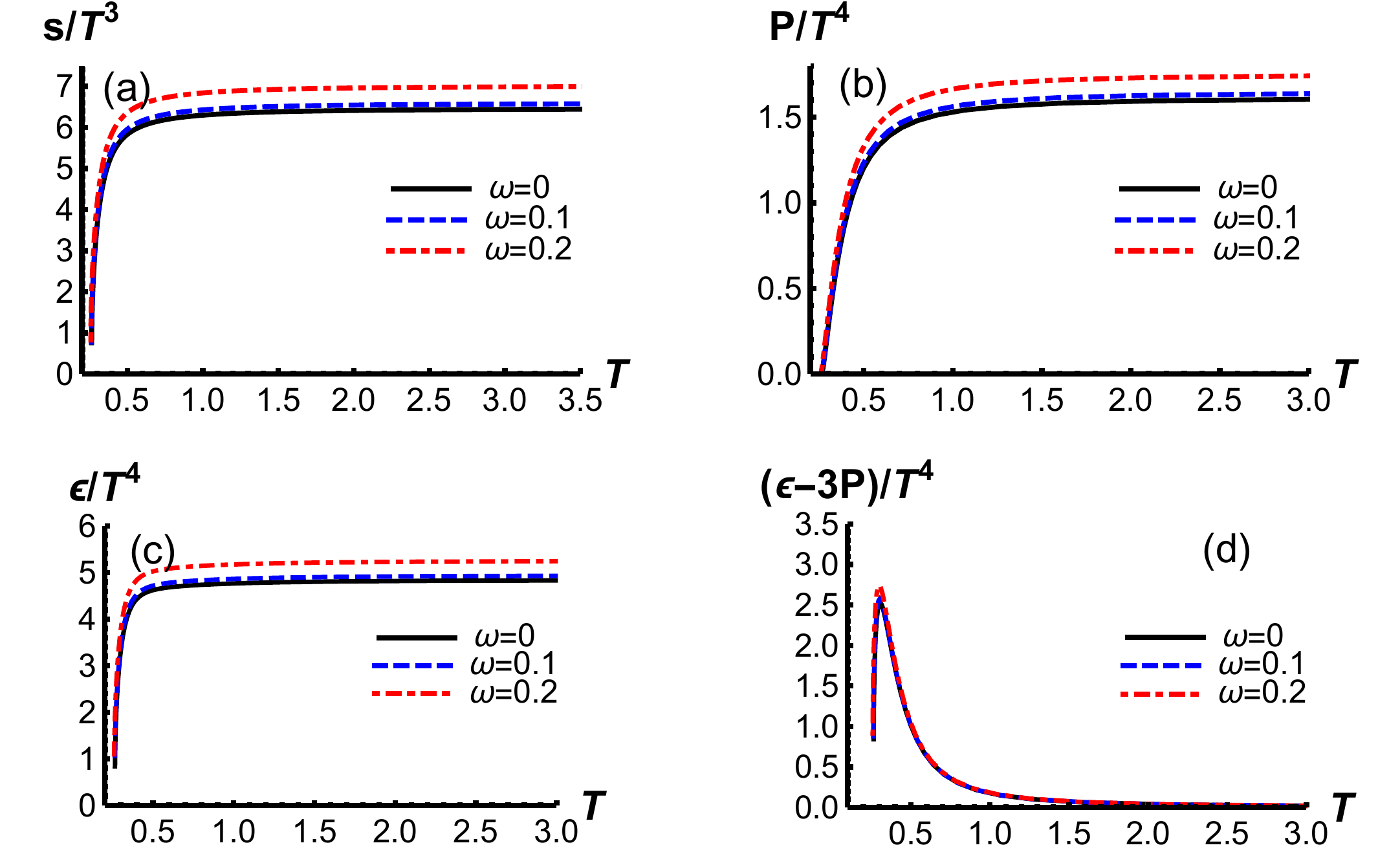}
	\caption{\label{eos} Thermodynamic properties of pure gluon system for different angular velocities:Entropy(a), pressure(b), energy(c) and trace anomaly(d) as a function $T$ at vanishing chemical potential. Black solid line is $\omega = 0\rm{GeV}$, blue dashed line is $\omega = 0.1\rm{GeV}$ and red dot-dashed line is $\omega = 0.2\rm{GeV}$. The unit of $T,\omega$ is GeV.}
\end{figure}

\begin{figure}
	\centering
	\includegraphics[width=15cm]{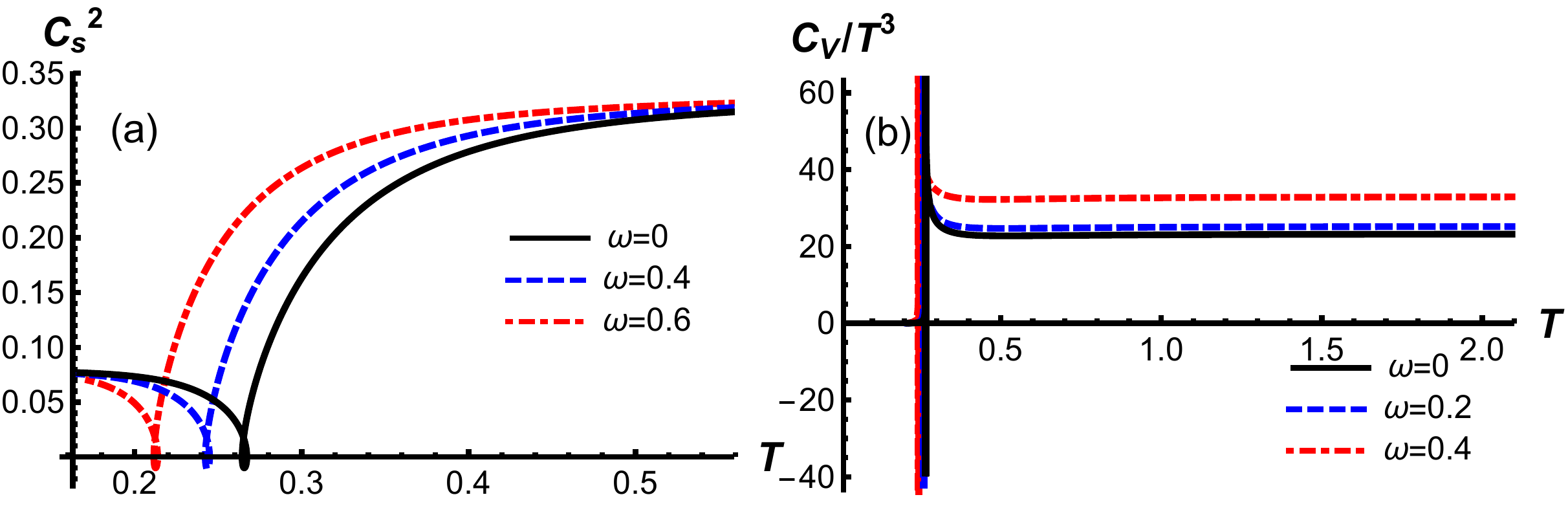}
	\caption{\label{cscv} For pure gluon system, the squared speed of sound as a function of $T$ at vanishing chemical potential. (a) for different angular velocities $\omega = 0$, $\omega = 0.4\rm{GeV}$ and $\omega = 0.6\rm{GeV}$ and the specific heat(b) as a function of T at vanishing chemical potential of pure gluon system for different angular velocities $\omega = 0$, $\omega = 0.2\rm{GeV}$ and $\omega = 0.4\rm{GeV}$. The unit of $T,\omega$ is \rm{GeV}.}
\end{figure}

Then, the other thermodynamic quantities under rotation can be computed in the same way as in two-flavor system. In Fig.\ref{eosall}, we present our model calculation without rotation with the blue solid line. We can see that the results agree very well with the blue dots from lattice simulation\cite{Boyd:1996bx}.  The results for $\omega=0.1,0.2\rm{GeV}$ are shown in Fig.\ref{eos}. We can find that the behavior of EoS under rotation for pure gluon system is similar to the two-flavor case. The negative part of the specific heat and square of sound velocity correspond to the thermodynamic instability in Fig.\ref{cscv}.

\subsection{Deconfinement phase transition under rotation}
\begin{figure}
	\centering
	\includegraphics[width=15cm]{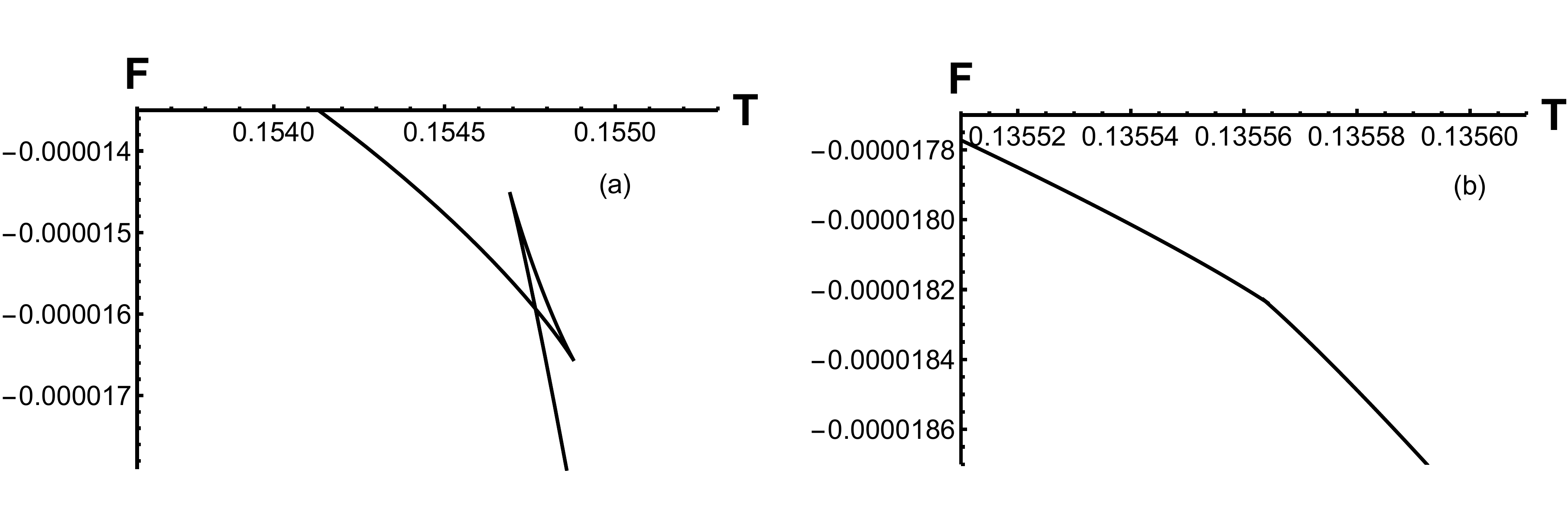}
	\caption{\label{freeenergy} For pure gluon system, the free energy as a function of $T$ at chemical potential $\mu = 0.1GeV$. (a)Angular velocities $\omega = 0.6GeV$. (b)Angular velocities $\omega = 0.67GeV$. The unit of $T,F$ is \rm{GeV}.}
\end{figure}

From the above study, we could see that the phase transition for two-flavor system, as well as pure gluon system with finite $\mu$ and large $\omega$ is crossover. We define the pseudo-transition temperature where $C_s^2$ reaches its minimum.  For pure gluon system with small $\omega$, the transition is a first order one and we will define the transition point where the free energy of the low temperature branches and high temperature branches intersects in Fig.\ref{freeenergy}. Under this definition, we could extract the transition temperature as a function of $\mu$ and $\omega$, and obtain the phase diagram in $T-\mu$ and $T-\omega$ plane.

\begin{figure}
	\centering
	\includegraphics[width=15cm]{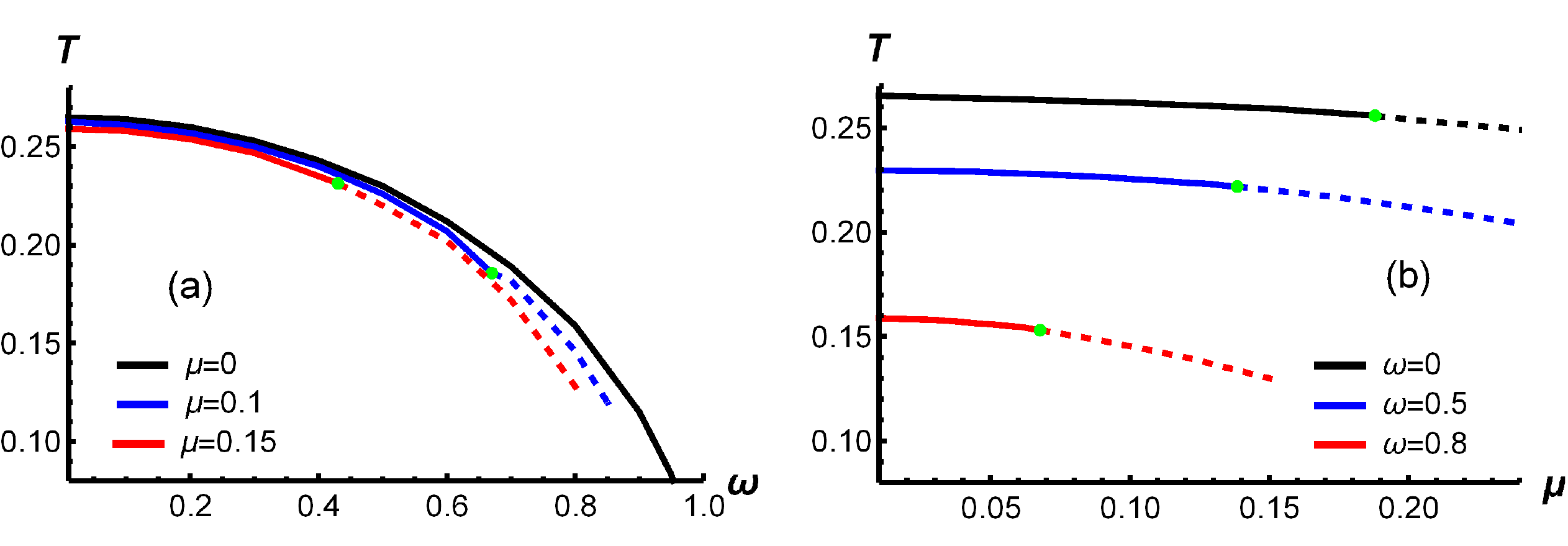}
	\caption{\label{purephase} Deconfinement phase diagram for pure gluon system in the $T- \omega$ and $T- \mu$ plane. Solid line is first order transition and green point is the CEP. (a) Deconfinement phase diagram in the $T - \omega$ plane for $\mu = 0, 0.1, 0.15\rm{GeV}$. Positions of CEP are located at $(\omega^E,T^E)=(0.67, 0.186), (0.43, 0.231)$ respectively. (b) Deconfiement phase diagram in the $T - \mu$ plane for $\omega = 0, 0.5, 0.8\rm{GeV}$. Positions of CEP are located at $(\mu^E,T^E)=(0.188, 0.256), (0.139, 0.222), (0.068, 0.153)$ respectively. The unit of $T,\mu,\omega$ is GeV.}
\end{figure}

\begin{figure}
	\centering
	\includegraphics[width=15cm]{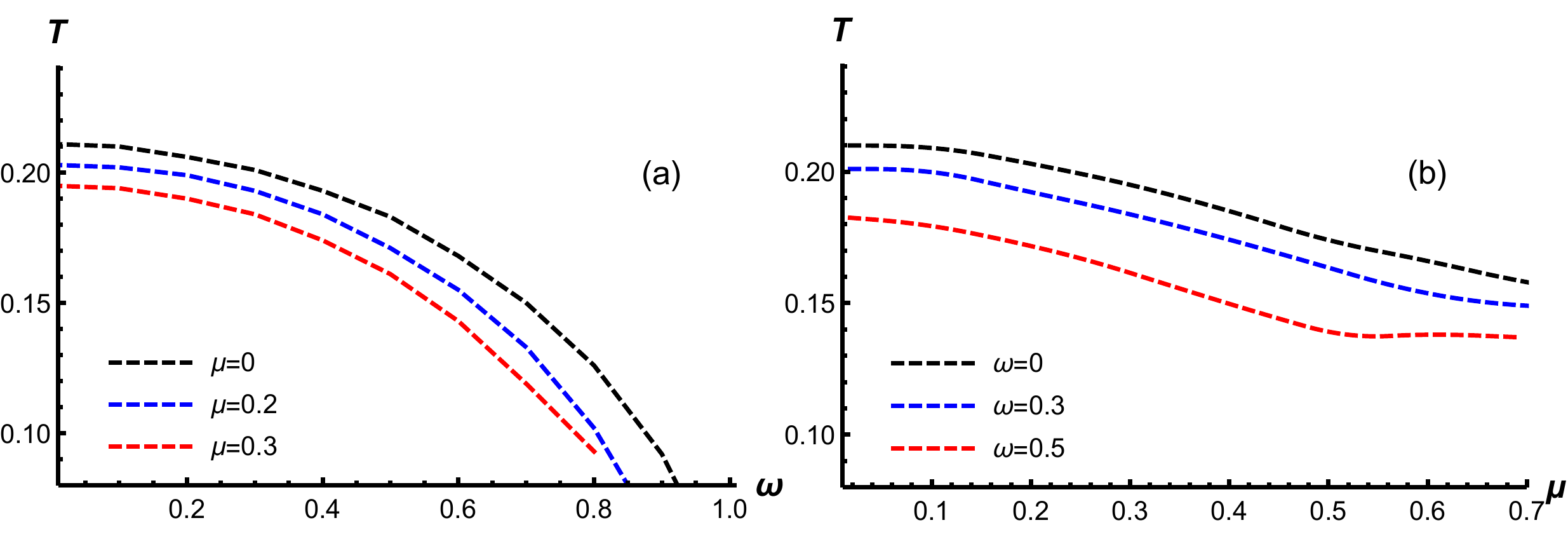}
	\caption{\label{lightphase} Deconfinement phase diagram for quenched two-flavor system in the $T- \omega$ and $T-\mu$ plane. (a) Deconfinement phase diagram for quenched two- flavor system in the $T- \omega$  plane for $\mu = 0$(black dashed line), $\mu = 0.2\rm{GeV}$(blue dashed line) and $\mu = 0.3\rm{GeV}$ (red dashed line). (b) Deconfinement phase diagram for quenched two-flavor system in the $T- \mu$  plane for $\omega = 0$ (black dashed line), $\omega = 0.3\rm{GeV}$ (blue dashed line) and $\omega = 0.5\rm{GeV}$ (red dashed line). The unit of $T,\mu,\omega$ is GeV.}
\end{figure}

We show the phase diagram of deconfinement for pure gluon system and two-flavor system under rotation in $T - \mu$ and $T - \omega$ plane in Fig.\ref{purephase} and \ref{lightphase}, respectively. It is shown that  $T_c=0.265\rm{GeV}$ for pure gluon system and $T_c=0.211\rm{GeV}$ for two-flavor system when $\mu=0,\omega=0$. Fig.\ref{purephase} (a) shows the  deconfinement phase transition for pure gluon system in the $T - \omega$ plane with different chemical potentials. At vanishing chemical potential $\mu=0$, the deconfinement transition is of first order phase transition in the whole $T - \omega$ plane. With the increase of chemical potential, the critical end point(CEP) shows up. Different from the chiral phase transition case shown in \cite{Wang:2018sur}, the deconfinement phase transition is of 1st order at small angular velocity and of crossover at large angular velocity.  When $\mu$ increases, the phase transition line is almost the same, only the CEP shifts to smaller angular velocity along the phase transition line. Fig.\ref{purephase} (b) shows the deconfinement phase transition in the $T - \mu$ plane for different angular velocities $\omega$. When $\omega=0$, the deconfinement phase transition is of 1st order phase transition in lower chemical potential and of crossover at higher chemical potential, and the CEP is located at $(\mu^E,T^E)=(0.188, 0.256) {\rm GeV}$. When the angular velocity increases, the phase transition line shifts down, and the location of the CEP shifts to the lower left plane.

The phase diagram of two-flavor system has be shown in Fig.\ref{lightphase}. We can see that the angular velocity and chemical potential will suppress the transition temperature and the phase transition will be always crossover in the whole phase diagram. Since the angular velocity is normalized, the transition temperature will decrease down to zero at $\omega \rightarrow 1GeV$. In the left panel of Fig.\ref{lightphase}, we can see that small angular velocity has less influence on transition temperature while large angular velocity leads to a quick decrease of phase transition temperature. The right panel of Fig.\ref{lightphase} shows that the phase transition temperature has weak dependence of chemical potential, which is similar to our previous work and PNJL model of deconfinement transition of light flavor\cite{Chen:2019rez,Abuki:2008nm,McLerran:2008ua}. Besides, rotation will lead to a anisotropic background which is in parallel with the analysis in anisotropic theories\cite{Giataganas:2012zy,Giataganas:2017koz}. There it has been found that the presence of anisotropy leads to easier dissociation of the
Q$\rm{\bar{Q}}$ and that in phase transitions anisotropy acts like a catalyst decreasing the critical temperature.\footnote{Thanks Dimitrios Giataganas for this commment.}

\subsection{Heavy-quark potential, Polyakov loop and spatial Wilson loop under rotation}
The above phase transition structure is obtained by analyzing the geometric phase transition, i.e. extracted from the thermodynamic quantities. To get more information of the transition, we go further to investigate the order parameters, which reveals the symmetry broken and restored in the transition. Since usually, the geometric phase transition would be considered as deconfinement phase transition, we will examine the variation of the loop operators relevant quantities in this section. Usually, the expectation value of Polyakov loop is considered as the order parameter of deconfinement phase transition, and the heavy quark potential and spatial Wilson loop reflect the color-electric and color-magnetic glue-dynamics, which are also strongly correlated with deconfinement transition. Therefore, in this section, we will study the rotating effect on those quantities.

\begin{figure}
	\centering
	\includegraphics[width=15cm]{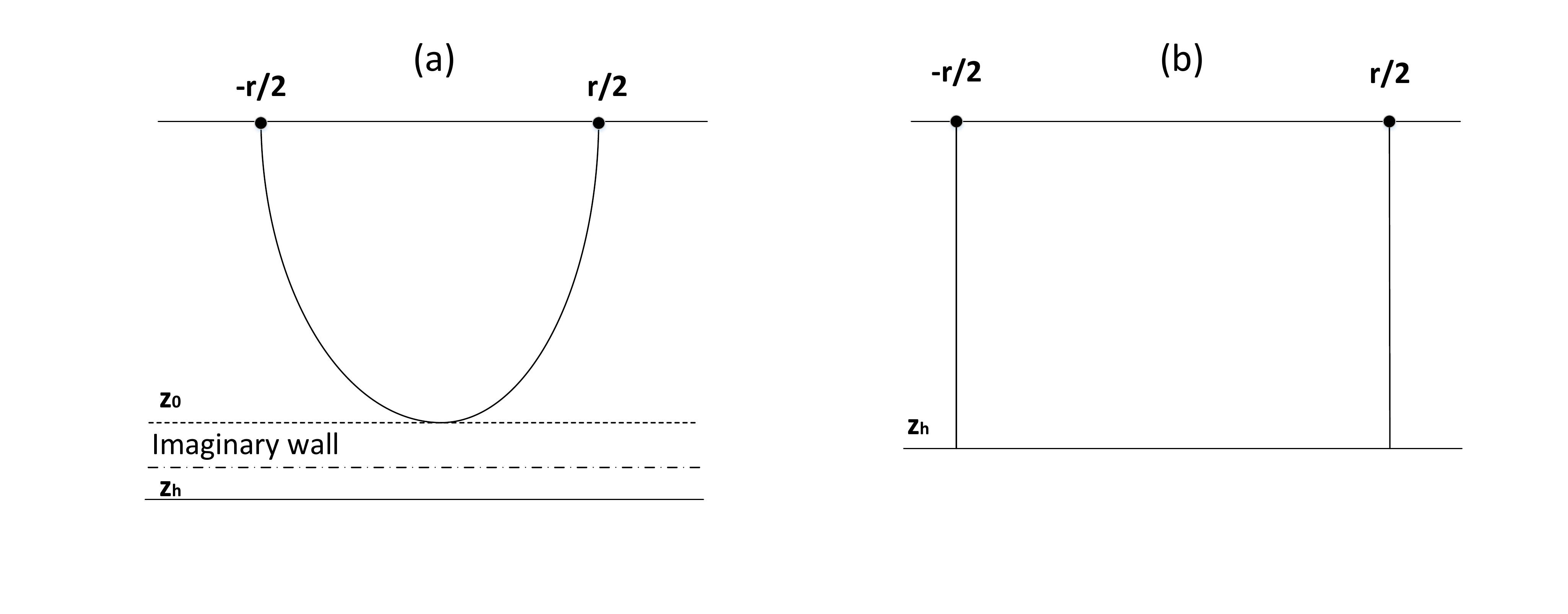}
	\caption{\label{quarkonium} The quark-antiquark pair locating at $\frac{-r}{2}$ and $\frac{r}{2}$ is connected by a U-shaped string. $z_0$ is the vertex of U-shape string. $z_h$ is the position of black hole horizon. (a)specious-confined phase. (b) Deconfined phase. }
\end{figure}

\begin{figure}
	\centering
	\includegraphics[width=15cm]{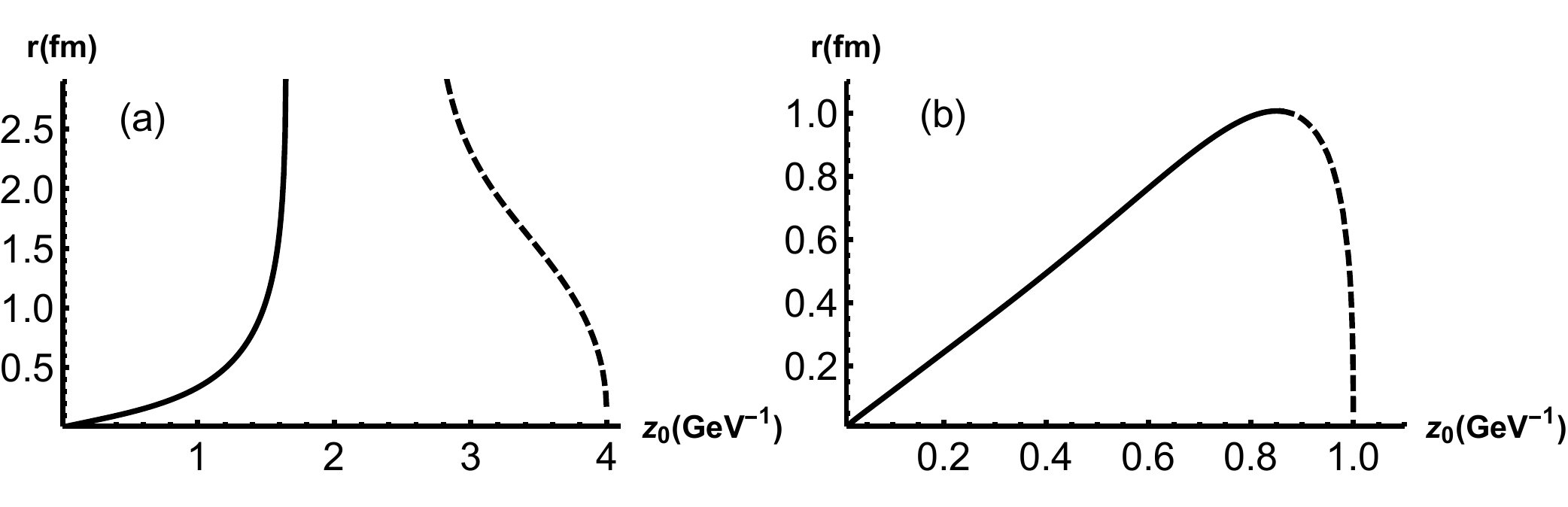}
	\caption{\label{rz0} The quark-antiquark distance r as a function of $z_0$. (a) specious-confined phase for $z_h = 4GeV^{-1}$. (b) Deconfined phase for $z_h = 1{\rm GeV}^{-1}$.}
\end{figure}

In Fig.\ref{quarkonium}, we only consider a heavy-quark probe in the specious-confinement and deconfiment phase for different temperatures. In the specious-confined phase, There is a imaginary wall, the string can't go beyond the wall with the increase of separate distance. The quark-antiquark pair always is connected by a U-shaped string. In deconfiment phase, there exists a maximum inter-quark distance beyond which the quark-antiquark pair will dissolve. To be more clear, we draw the figure of inter-quark distance v.s. $z_0$ with different temperatures at specious-confinement/deconfinement phase in Fig.\ref{rz0}. When we put a heavy-quark probe into the system, we can see that the physical inter-quark distance represented by solid line can go to infinity at low temperature(specious-confinement phase) and quark-antiquark will dissolve at a maximum distance(deconfinement phase). In the deconfinement phase, the U-shaped string will become two straight strings at a certain distance. Nevertheless, the U-shaped string will always exist in the specious-confinement phase. We can also refer to Ref.\cite{Dudal:2017max,Andreev:2006nw} about more analysis of heavy-quark free energy in the specious-confinement and deconfinement phase.

According to holographic dictionary, the loop operators could be extracted by minimizing the area of string world-sheet, which is related to the string motion. Thus, we transfer the metric to string frame from Einstein frame by setting $A_s(z) = A_e(z) + \sqrt{\frac{1}{6}} \phi(z)$, see \cite{Li:2011hp,Dudal:2017max} for details. As a concequence, the static metric in string frame is

\begin{equation}d s^{2}=\frac{e^{2 A_s(z)}}{z^{2}}\left[-G(z) d t^{2}+\frac{1}{G(z)} d z^{2}+d \vec{x}^{2}\right].\end{equation}

The rotating background can be similarly written in the same way as before.

 \begin{equation}
ds^2 = -N_1(z)dt^2 + \frac{H_1(z)dz^2}{G(z)} + R_1(z) ( d\theta + P(z)dt)^2 + H_1(z)\sum\limits_{i=1}^2dx_i^2,
\end{equation}
with
\begin{align}
N_1(z) &= \frac{H_1(z)G(z)(1-\omega^2 l^2)}{1-G(z)\omega^2 l^2}, \\
H_1(z) &= \frac{L^2e^{2A_s(z)}}{z^2}, \\
R_1(z) &= H_1(z)\gamma^2 l^2 - H_1(z)G(z)\gamma^2\omega^2 l^4,  \\
P(z) &= \frac{\omega - G(z)\omega }{1 - G(z)\omega^2 l^2},\\
\gamma &= \frac{1}{\sqrt{1-\omega^2l^2}}.
\end{align}

 Then, we put a static heavy quark-antiquark pair in the rotating background to probe the phase transition. The string world-sheet action is defined by the Nambu-Goto action and takes the following form

\begin{equation}S_{N G}=-\frac{L^2}{2 \pi \alpha^{\prime}} \int \mathrm{d}^{2} \xi \sqrt{-\operatorname{det} g_{a b}}.\end{equation}

Here, $g_{a b}$ is the induced metric defined as
\begin{equation}g_{a b}=g^s_{M N} \partial_{a} X^{M} \partial_{b} X^{N}, \quad a, b=0,1,
\end{equation}
and $\alpha'$ is the string tension. Here, $X^{M}$ and $g^s_{M N}$ are the coordinates and the string-frame metric, respectively.

To calculate the quark-antiquark potential, we consider the string ends at a static quark-antiquark pair locating at $x_1=-r/2$ and $x_1=r/2$. A simplest parametrization of the string world-sheet parameters is $\xi^{0}=t, \xi^{1}=x_1$. Under this condition, the effective Nambu-Goto action can be written as
\begin{equation}
S_{N G}=-\frac{L^2}{2 \pi \alpha^{\prime} T} \int_{-r / 2}^{r / 2} \mathrm{d} x_1 \sqrt{k_{1}(z) \frac{\mathrm{d} z^{2}}{\mathrm{d} x_1^{2}}+k_{2}(z)}.
\end{equation}

Here,  we define
\begin{equation}\begin{aligned}
		k_{1}(z) &=\frac{(R_1(z)-N_1(z)) H_1(z)}{G(z)}, \\
		k_{2}(z) &=-(R_1(z)-N_1(z)) H_1(z).
\end{aligned}\end{equation}
 The expectation value of the Wigner-Wilson loop is then related to the on-shell string action by

\begin{equation}\langle W(\mathcal{C})\rangle=\int D X e^{-S_{N G}} \simeq e^{-S_{on-shell}},\end{equation}
where $\mathcal{C}$ denotes a closed loop in spacetime. The definition of the heavy-quark potential is\cite{Maldacena:1998im,Rey:1998ik,Ewerz:2016zsx}

\begin{equation}\langle W(\mathcal{C})\rangle \sim e^{-V(r, T) / T},\end{equation}
where $r$ is the separate distance of quarks. So, to get the potential, one has to solve the on-shell string world-sheet action.

To do that, following the standard procedure\cite{Li:2011hp,Andreev:2006nw,Colangelo:2010pe,Yang:2015aia,Chen:2017lsf}, we can define an effecive `Hamitonian'
\begin{equation}
\mathcal{H} = z' \frac{\partial{\mathcal{L}}}{z'} - \mathcal{L} = \frac{k_2(z)}{\sqrt{k_1(z) z'^2 + k_2(z)}}.\end{equation}
Solving  $z'$ form the equation
\begin{equation} \frac{k_2(z)}{\sqrt{k_1(z) z'^2 + k_2(z)}} = \frac{k_2(z_0)}{\sqrt{k_2(z_0)}}, \end{equation}
we can obtain the inter-quark distance and renormalized potential of heavy quark-antiquark pair as
\begin{equation}r=\int_{-\frac{r}{2}}^{\frac{r}{2}} d x=2 \int_{0}^{z_{0}} d z \frac{1}{z^{\prime}}= 2 \int_{0}^{z_{0}}\left[\frac{k_{2}(z)}{k_{1}(z)}\left(\frac{k_{2}(z)}{k_{2}\left(z_{0}\right)}-1\right)\right]^{-1 / 2} \mathrm{d} z,\end{equation}

\begin{equation}\frac{ V }{\sqrt{\lambda}}=\frac{1}{\pi}(\int_{0}^{z_{0}} \mathrm{d} z(\sqrt{\frac{k_{2}(z) k_{1}(z)}{k_{2}(z)-k_{2}\left(z_{0}\right)}}-\sqrt{k_{2}(z \rightarrow 0, \omega \rightarrow 0)})-\int_{z_{0}}^{\infty} \sqrt{k_{2}(z \rightarrow 0, \omega \rightarrow 0)} \mathrm{d} z).\end{equation}

\begin{figure}
	\centering
	\includegraphics[width=15cm]{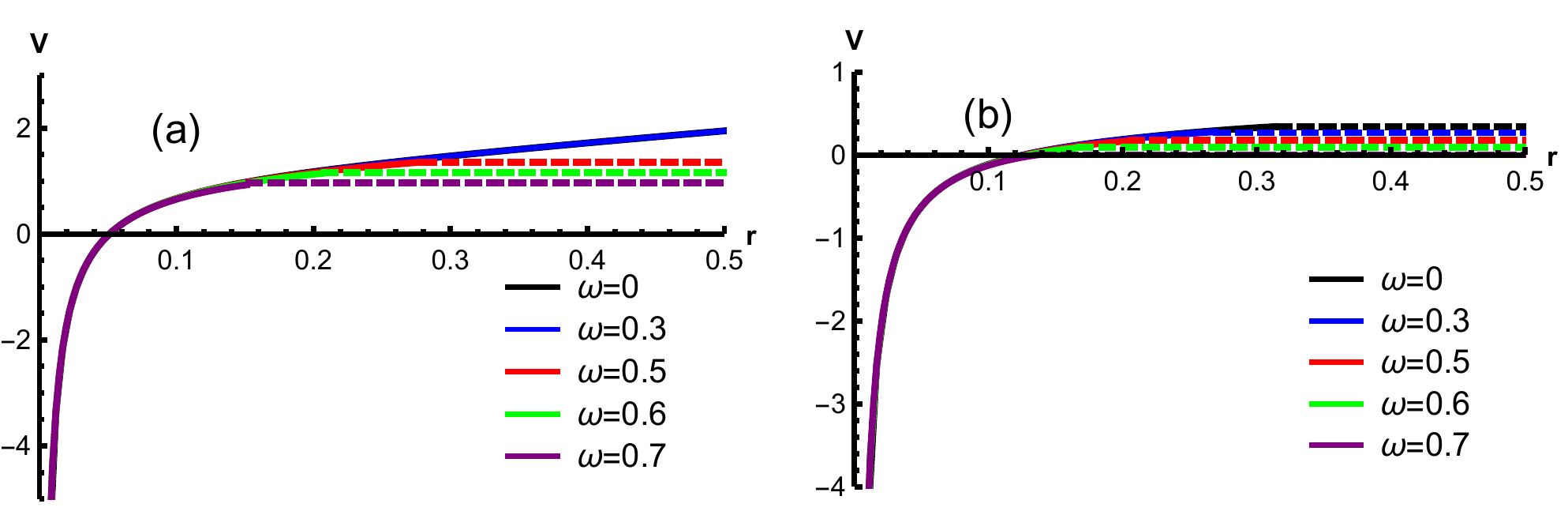}
	\caption{\label{potential} The potential of quark-antiquark pair as a function of separate distance at fixed $T = 0.25\rm{GeV}$ and $\mu$ = 0 for different angular velocities $\omega = 0$(black line),  $\omega = 0.3\rm{GeV}$(blue line),  $\omega = 0.5\rm{GeV}$(red line),   $\omega = 0.6\rm{GeV}$(green line) and $\omega = 0.7\rm{GeV}$(purple line). The unit of T is in GeV and r in fm. (a) In the pure gluon system. (b) In the two-flavor system. }
\end{figure}

Here, $z_0$ is the vertex of U-shape string as shown in Fig.\ref{quarkonium}. The effective coupling $\sqrt{\lambda}\equiv\frac{L^{2}}{\alpha^{\prime}}$ will be set it to 1, since we only focus on the qualitative behavior. In the above expression, $\sqrt{k_{2}(z \rightarrow 0, \omega \rightarrow 0)}$ is a counter term to cancel the divergence near $z=0$ and can be calculated as $\frac{1}{z^2}$. The potential of heavy quark-antiquark pair as a function of $r$ is shown in Fig.{\ref{potential}}. From the figure, we could see that for pure gluon system, at $T=0.25\rm{GeV}$ and $\omega=0$, there is linear part in $V(r,T)$, showing confinement of heavy quarks. Actually, since we are considering the time-like loop, the potential is related to the color-electric modes. Then, when we increase $\omega$, we could see that the linear part would be screened, which might be considered as a signal of the deconfining quarks. As for two-flavor system, at $T=0.25\rm{GeV}$, the linear part of quark-antiquark potential has already been screened and $\omega$ would enhance the screening effect. Therefore, we could see that color-electric modes are confined at small angular velocity and at large angular velocity they would deconfine.

\begin{figure}
	\centering
	\includegraphics[width=15cm]{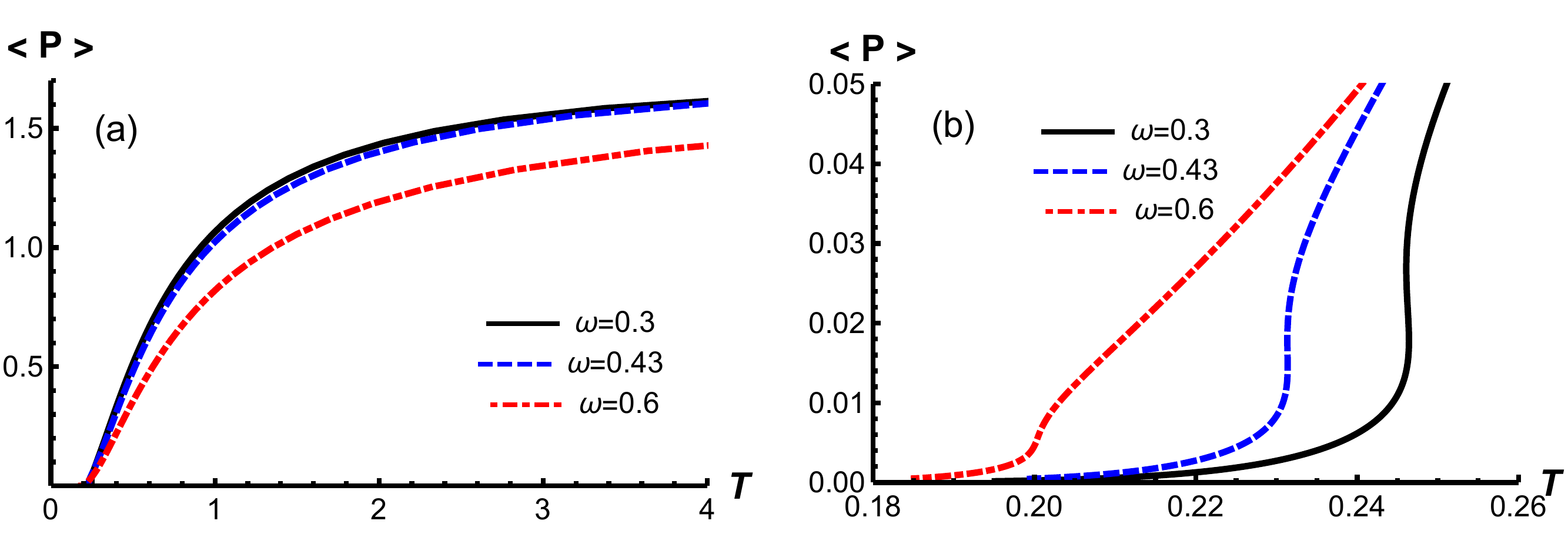}
	\caption{\label{polyakov} (a)In pure gluon system, the expectation value of a single Polyakov loop as a function of T at $\mu = 0.15\rm{GeV}$ for different angular velocities of $\omega = 0.3$(solid black line), $\omega = 0.43\rm{GeV}$(dashed blue line) and $\omega = 0.6\rm{GeV}$(dot-dashed red line). (b) An enlarged view of (a).The unit for $T,\mu$ is in GeV.}
\end{figure}

\begin{figure}
	\centering
	\includegraphics[width=15cm]{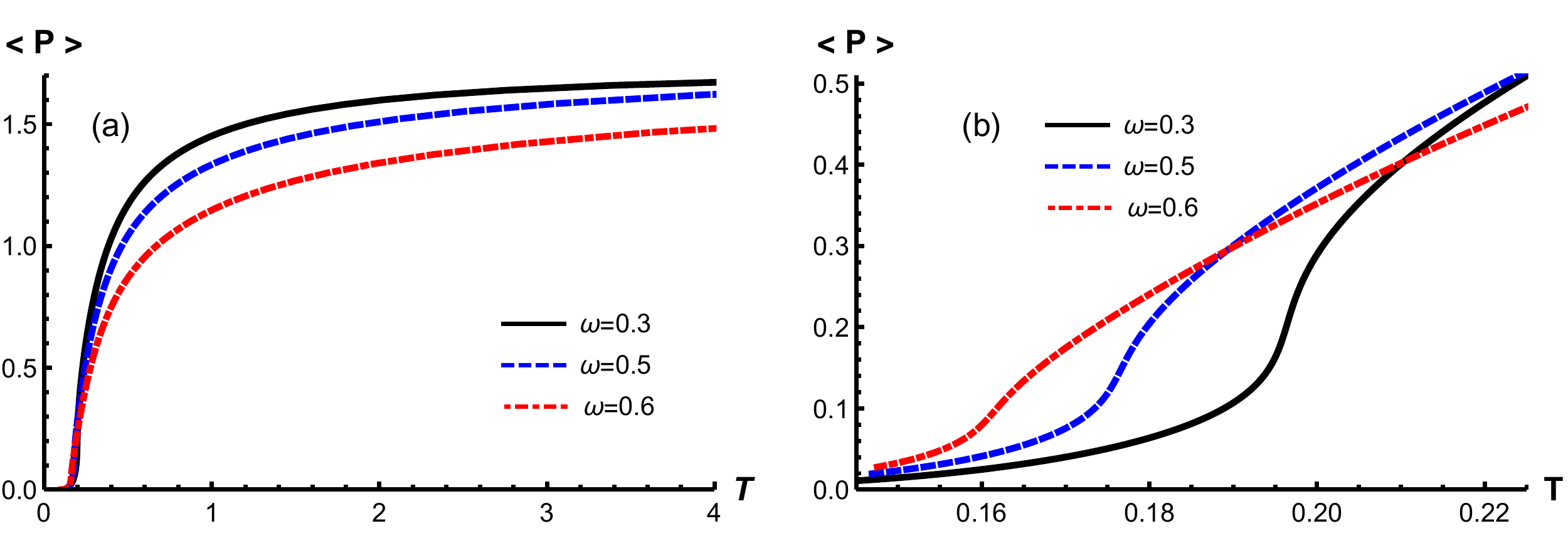}
	\caption{\label{polyakov2} (a) In two-flavor system, the expectation value of a single Polyakov loop as a function of T at $\mu = 0.15\rm{GeV}$ for different angular velocities of $\omega = 0$(solid black line), $\omega = 0.5\rm{GeV}$(dashed blue line) and $\omega = 0.6\rm{GeV}$(dot-dashed red line). (b) An enlarged view of (a).The unit for $T,\mu$ is in GeV.}
\end{figure}

Then, we will move to the order parameter of deconfinement transition, i.e. the expectation value of Polyakov loops\cite{Polyakov:1978vu}. Following \cite{Colangelo:2010pe}, at finite temperature, the correlation function of two Polyakov loops can be obtained in QCD from the free energy(or potential) $V(r, T)$ of an infinitely heavy quark-antiquark pair at distance $r$:

\begin{equation}\left\langle\mathcal{P}\left(\vec{x}_{1}\right) \mathcal{P}^{\dagger}\left(\vec{x}_{2}\right)\right\rangle=e^{-\frac{1}{T} V(r, T)+\gamma(T)}
\end{equation}
with $r=\left|\vec{x}_{1}-\vec{x}_{2}\right|$ and $\gamma(T)$ a normalization constant. The expectation value of a single Polyakov loop
\begin{equation}
\langle\mathcal{P}\rangle=e^{-\frac{1}{2 T} V(r=\infty, T)}.
\end{equation}
Thus, from the long distance behavior of heavy-quark potential $V(r,T)$, one can get $\langle P\rangle$. If the quarks are confined, $V(\infty, T)=+\infty$, and $\langle P \rangle=0$ for a pure SU(N) theory, while it is finite in deconfinement phase. We obtain the numerical results of $\langle P \rangle$, as shown in the left panels of Fig.\ref{polyakov} and Fig.\ref{polyakov2}.  From the figures, we can see that both for the two kind of systems, at low temperature, $\langle P\rangle$ is rather small, showing the confinement of color degrees. At high temperature,  $\langle P\rangle$ would increase to a finite constant, showing the deconfinement of color degrees. Therefore, we could see that for both the two systems, the geometric phase transition is really a description of deconfinement phase transition.

Furthermore, we could see from the order parameter that the rotation effect would affect the transition a lot. Both for the two systems, the constant at large $T$ limit will be enhanced with the increase of angular velocity, which again shows enhancement effect on the deconfinement from rotation. For the transition order and the location of transition point, the two systems are different.  We zoom in the area near the transition point in the right panels of Fig.\ref{polyakov} and Fig.\ref{polyakov2}. It is shown that for pure gluon system, the expectation value of a single Polyakov loop changes from multiple value to single value with the increase of angular velocity, indicating the transition from first order transition to crossover. However, for two-flavor system, the Polyakov loop is always single valued which indicates the phase transition is always crossover. For the crossover transition part, we can define the pseudo-transition temperature where $\langle P\rangle $ changes fastest, i.e. the location of maximum $|\frac{d\langle P\rangle}{dT}|$. Then, we find that the crossover line are very close to the one obtain from thermodynamic quantities. For the first order phase transition, we still extract the phase transition temperature from the free energy. There is a cross point of the free energy which is defined as the first order transition point\cite{Dudal:2017max}.

Finally, we will consider the color-magnetic sector, i.e. consider the loop in a space-like contour, and investigate the spatial Wilson loop. From the dictionary, the spatial Wilson loop can be calculated similarly but choosing a different way of parameterization, due to the space-like property of the loop. Here, we can choose the string world-sheet coordinates as $\xi^{0}=x_1, \xi^{1}=l\theta$. Then, similar as the heavy quark potential, we get the spatial potential as
\begin{equation}\frac{V_s}{\sqrt{\lambda}}=\frac{1}{\pi}\left(\int_{0}^{z_{0}} \mathrm{d} z(\sqrt{\frac{K_{2}(z) K_{1}(z)}{K_{2}(z)-K_{2}\left(z_{0}\right)}}-\sqrt{K_{2}(z \rightarrow 0, \omega \rightarrow 0)})-\int_{z_{0}}^{\infty} \sqrt{K_{2}(z \rightarrow 0, \omega \rightarrow 0)} \mathrm{d} z\right)\end{equation}
with $K_{1}(z) = \frac{R_1(z)H(z)}{G(z)}, K_2(z) = R_1(z)H(z)$.

\begin{figure}
	\centering
	\includegraphics[width=15cm]{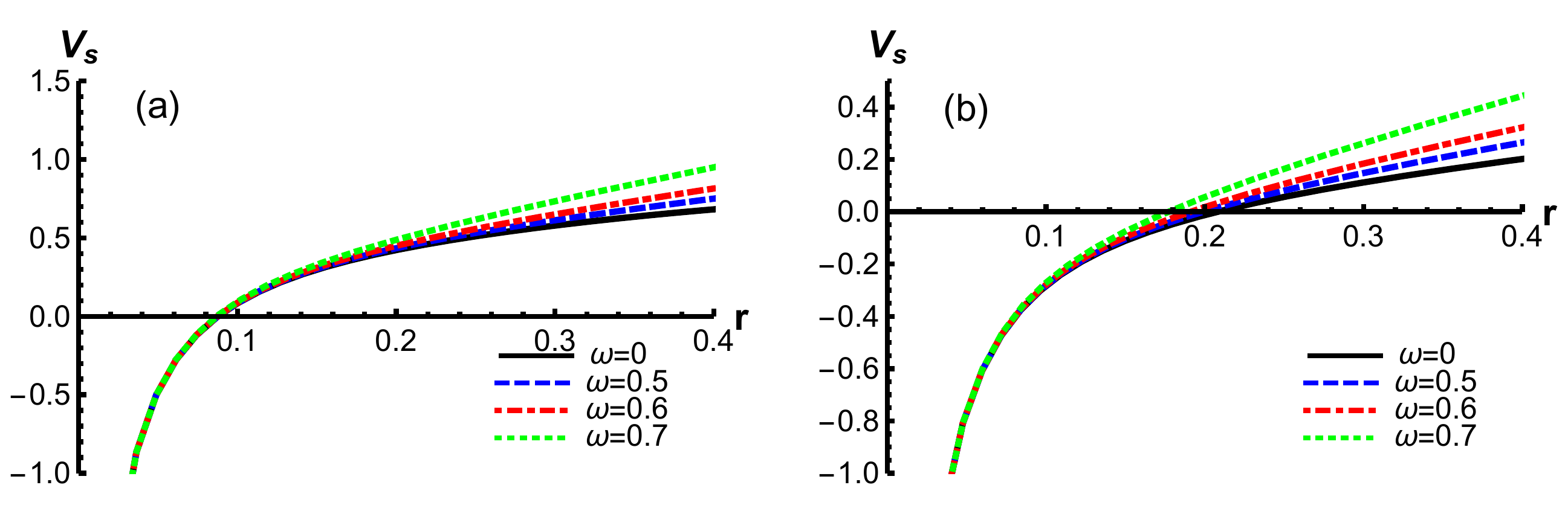}
	\caption{\label{Spotential} The spatial Wilson loop as a function of separate distance r at $\mu = 0\rm{GeV}$ and $T = 0.25\rm{GeV}$ for different anugular velocities of $\omega = 0$,  $\omega = 0.5\rm{GeV}$, $\omega = 0.6\rm{GeV}$ and $\omega = 0.7\rm{GeV}$. The unit of $T,\mu,\omega$ is GeV and $r$ is in $\rm{fm}$. (a) Pure gluon system. (b) Two-flavor system.}
\end{figure}

Inserting the string frame metric, one can get the numerical results of the spatial potential, as shown in Fig.\ref{Spotential}. From the figure,  it can be seen that for both the two systems, there are linear part at large distance in the spatial potential. The slope of the potential would be enhanced by the angular velocity. The effect of angular velocity on the spatial Wilson loop is similar to the temperature as shown in Lattice\cite{Bali:1993tz}. The enhancement in the spatial potential shows the confinement of color-magnetic modes, which is different from the color-electric modes.

\label{sec:quenchedgluon}

\section{Conclusion}
\label{sec-sum}
We investigate the rotation effect to deconfinement phase transition in a holographic QCD model, which is an simple extension of the EMD model in Ref.\cite{Dudal:2017max} to finite angular velocity case. By doing the local Lorentz boost, we obtain the 5D dual geometric background under rotation and consider it as an approximate dual description of the 4D rotating QCD matter.

Firstly, we fix the model parameters by analyzing thermodynamic quantities without rotation. We identify the Hawking temperature and Beikenstein-Hawking entropy as the temperature and entropy of the 4D system. Then, from the thermodynamic relations we could extract the pressure, energy density, trace anomaly, sound speed and specific heat. By comparing the results to lattice simulations without rotation, we fix all the model parameters. It is found that such a model could describe thermodynamics of both pure gluon system and two-flavor system without rotation well. For pure gluon system, there is a first order phase transition at around $T_c=0.265\rm{GeV}$, while for two-flavor system it is a first order transition locating at $T_c=0.211\rm{GeV}$.

Then, we consider QCD matter with finite baryon number density and under rotation. It is shown that the $T$ scaled entropy density, pressure, energy density and specific heat will be enhanced by angular velocity. From the thermodynamic quantities, we could see that the location of the phase transition varies with chemical potential $\mu$ and angular velocity $\omega$. For pure gluon system, the transition is always a first order one when $\mu=0$. At finite $\mu$, for a fixed $\omega$, the transition temperature would decrease with the increase of $\mu$. But the decreasing rate is quite low, within $2.2\%$ decreasing from $\mu=0$ to $\mu=0.15\rm{GeV}$. The transition temperature would decrease with $\omega$ as well. Furthermore, for finite $\mu$,  the transition would turn to a crossover one at large $\omega$. In between the first order line and crossover region, there is a critical end point. The location of the CEP would shift towards low temperature and small density with the increasing of $\omega$. As for two-flavor case, it is shown that the transition is always of crossover type. With the increasing of $\omega$, the transition temperature would decrease very fast to zero at $\omega=1\rm{GeV}$. For a fixed $\omega$, the transition temperature would decrease with $\mu$, but in a lower rate.

To get full understanding of the phase transition, and to check the physical contents of the geometric phase transition, we investigate the order parameter of deconfinement transition, i.e. the loop operators. By extracting the expectation value of Polyakov loop, we could see that both for pure gluon system and two-flavor system, around the phase transition point extracting from thermodynamic quantities, the Polyakov loop would undergo a jump or fast increase. This confirms that the transition is a deconfinement transition. Furthermore, one could see that the rotation effect would cause an enhanced screen effect on quark-antiquark potential. For the spatial potential, which is controlled by color magnetic field, the slope of the linear part increase with angular velocity. From this results, it seems that the coupling of angular velocity with color electric field and color magnetic are different. But the details require further study, and we will leave them to the future. In this work, we only consider deconfinement phase transition, and the results are just a preliminary approximation to the full solution. It would be quite interesting to study chiral phase transition together with deconfinement phase transition in a full HQCD model.

\vskip 1.5cm
{\bf Acknowledgement}
\vskip 0.2cm
D.L. is supported by the NSFC under Grant Nos. 11805084 and 11647141, D.F.H is supported  by the NSFC under Grant Nos. 11735007 and 11875178, and M.H. is supported in part by the NSFC under Grant Nos. 11725523 and 11735007,  Chinese Academy of Sciences under Grant No. XDPB09, the start-up funding from University of Chinese Academy of Sciences(UCAS), and the Fundamental Research Funds for the Central Universities.

\vspace{10mm}
\vspace{-1mm}
\centerline{\rule{80mm}{0.1pt}}

\clearpage

\vspace{2mm}



\begin{thebibliography}{90}

\bibitem{Aoki:2006we}
  Y.~Aoki, G.~Endrodi, Z.~Fodor, S.~D.~Katz and K.~K.~Szabo,
  ``The Order of the quantum chromodynamics transition predicted by the standard model of particle physics,''
  Nature {\bf 443} (2006) 675
  [hep-lat/0611014].

\bibitem{Ding:2015ona}
  H.~T.~Ding, F.~Karsch and S.~Mukherjee,
  ``Thermodynamics of strong-interaction matter from Lattice QCD,''
  Int.\ J.\ Mod.\ Phys.\ E {\bf 24} (2015) no.10,  1530007
  [arXiv:1504.05274 [hep-lat]].

\bibitem{Aggarwal:2010cw}
  M.~M.~Aggarwal {\it et al.} [STAR Collaboration],
  ``An Experimental Exploration of the QCD Phase Diagram: The Search for the Critical Point and the Onset of De-confinement,''
  arXiv:1007.2613 [nucl-ex].

\bibitem{Odyniec:2013aaa}
  G.~Odyniec,
  ``RHIC Beam Energy Scan Program: Phase I and II,''
  PoS CPOD {\bf 2013} (2013) 043.

\bibitem{Luo:2017faz}
  X.~Luo and N.~Xu,
  ``Search for the QCD Critical Point with Fluctuations of Conserved Quantities in Relativistic Heavy-Ion Collisions at RHIC : An Overview,''
  Nucl.\ Sci.\ Tech.\  {\bf 28} (2017) no.8,  112
  [arXiv:1701.02105 [nucl-ex]].

\bibitem{Skokov:2009qp}
  V.~Skokov, A.~Y.~Illarionov and V.~Toneev,
  ``Estimate of the magnetic field strength in heavy-ion collisions,''
  Int.\ J.\ Mod.\ Phys.\ A {\bf 24} (2009) 5925
  [arXiv:0907.1396 [nucl-th]].

\bibitem{Voronyuk:2011jd}
  V.~Voronyuk, V.~D.~Toneev, W.~Cassing, E.~L.~Bratkovskaya, V.~P.~Konchakovski and S.~A.~Voloshin,
  ``(Electro-)Magnetic field evolution in relativistic heavy-ion collisions,''
  Phys.\ Rev.\ C {\bf 83} (2011) 054911
  [arXiv:1103.4239 [nucl-th]].

\bibitem{Bzdak:2011yy}
  A.~Bzdak and V.~Skokov,
  ``Event-by-event fluctuations of magnetic and electric fields in heavy ion collisions,''
  Phys.\ Lett.\ B {\bf 710} (2012) 171
  [arXiv:1111.1949 [hep-ph]].

\bibitem{Deng:2012pc}
  W.~T.~Deng and X.~G.~Huang,
  ``Event-by-event generation of electromagnetic fields in heavy-ion collisions,''
  Phys.\ Rev.\ C {\bf 85} (2012) 044907
  [arXiv:1201.5108 [nucl-th]].



\bibitem{Jiang:2016woz}
 Y.~Jiang, Z.~W.~Lin and J.~Liao,
  ``Rotating quark-gluon plasma in relativistic heavy ion collisions,''
  Phys.\ Rev.\ C {\bf 94}, no. 4, 044910 (2016)
  Erratum: [Phys.\ Rev.\ C {\bf 95}, no. 4, 049904 (2017)]
  [arXiv:1602.06580 [hep-ph]].

\bibitem{Deng:2016gyh}
  W.~T.~Deng and X.~G.~Huang,
  ``Vorticity in Heavy-Ion Collisions,''
  Phys.\ Rev.\ C {\bf 93}, no. 6, 064907 (2016)
  [arXiv:1603.06117 [nucl-th]].

\bibitem{Kharzeev:2007tn}
  D.~Kharzeev and A.~Zhitnitsky,
  Nucl.\ Phys.\  A {\bf 797}, 67 (2007).
  [arXiv:0706.1026 [hep-ph]].

\bibitem{Son:2009tf}
  D.~T.~Son and P.~Surowka,
  Phys.\ Rev.\ Lett.\  {\bf 103}, 191601 (2009).
 [arXiv:0906.5044 [hep-th]].

\bibitem{Kharzeev:2010gr}
  D.~E.~Kharzeev and D.~T.~Son,
  Phys.\ Rev.\ Lett.\  {\bf 106}, 062301 (2011).
  [arXiv:1010.0038 [hep-ph]].

\bibitem{Rajagopal:2015roa}
K.~Rajagopal and A.~V.~Sadofyev,
``Chiral drag force,''
JHEP \textbf{10} (2015), 018
[arXiv:1505.07379 [hep-th]].


\bibitem{Jiang:2015cva}
  Y.~Jiang, X.~G.~Huang and J.~Liao,
  ``Chiral vortical wave and induced flavor charge transport in a rotating quark-gluon plasma,''
  Phys.\ Rev.\ D {\bf 92}, no. 7, 071501 (2015).
  [arXiv:1504.03201 [hep-ph]].

\bibitem{Jiang:2016wvv}
  Y.~Jiang and J.~Liao,
  ``Pairing Phase Transitions of Matter under Rotation,''
  Phys.\ Rev.\ Lett.\  {\bf 117}, no. 19, 192302 (2016)
  [arXiv:1606.03808 [hep-ph]].

\bibitem{Wang:2018sur}
  X.~Wang, M.~Wei, Z.~Li and M.~Huang,
  ``Quark matter under rotation in the NJL model with vector interaction,''
  Phys.\ Rev.\ D {\bf 99}, no. 1, 016018 (2019)
  [arXiv:1808.01931 [hep-ph]].

\bibitem{Chen:2015hfc}
  H.~L.~Chen, K.~Fukushima, X.~G.~Huang and K.~Mameda,
  ``Analogy between rotation and density for Dirac fermions in a magnetic field,''
  Phys.\ Rev.\ D {\bf 93}, no. 10, 104052 (2016)
  [arXiv:1512.08974 [hep-ph]].


\bibitem{Ebihara:2016fwa}
  S.~Ebihara, K.~Fukushima and K.~Mameda,
  ``Boundary effects and gapped dispersion in rotating fermionic matter,''
  Phys.\ Lett.\ B {\bf 764}, 94 (2017)
  [arXiv:1608.00336 [hep-ph]].

  \bibitem{Chernodub:2016kxh}
  M.~N.~Chernodub and S.~Gongyo,
  ``Interacting fermions in rotation: chiral symmetry restoration, moment of inertia and thermodynamics,''
  JHEP {\bf 1701}, 136 (2017)
  [arXiv:1611.02598 [hep-th]].

\bibitem{Chernodub:2017ref}
  M.~N.~Chernodub and S.~Gongyo,
  ``Effects of rotation and boundaries on chiral symmetry breaking of relativistic fermions,''
  Phys.\ Rev.\ D {\bf 95}, no. 9, 096006 (2017)
  [arXiv:1702.08266 [hep-th]].

\bibitem{McLerran:2007qj}
  L.~McLerran and R.~D.~Pisarski,
  ``Phases of cold, dense quarks at large N(c),''
  Nucl.\ Phys.\ A {\bf 796} (2007) 83
  [arXiv:0706.2191 [hep-ph]].

\bibitem{McLerran:2008ua}
  L.~McLerran, K.~Redlich and C.~Sasaki,
  ``Quarkyonic Matter and Chiral Symmetry Breaking,''
  Nucl.\ Phys.\ A {\bf 824} (2009) 86
  [arXiv:0812.3585 [hep-ph]].

\bibitem{Li:2018ygx}
  Z.~Li, K.~Xu, X.~Wang and M.~Huang,
  ``The kurtosis of net baryon number fluctuations from a realistic Polyakov–Nambu–Jona-Lasinio model along the experimental freeze-out line,''
  Eur.\ Phys.\ J.\ C {\bf 79} (2019) no.3,  245
  [arXiv:1801.09215 [hep-ph]].





\bibitem{Maldacena:1997re}
  J.~M.~Maldacena,
  Int.\ J.\ Theor.\ Phys.\  {\bf 38}, 1113 (1999)
  [Adv.\ Theor.\ Math.\ Phys.\  {\bf 2}, 231 (1998)]
  [hep-th/9711200].

\bibitem{Gubser:1998bc}
  S.~S.~Gubser, I.~R.~Klebanov and A.~M.~Polyakov,
  ``Gauge theory correlators from noncritical string theory,''
  Phys.\ Lett.\ B {\bf 428}, 105 (1998)
  [hep-th/9802109].

\bibitem{Witten:1998qj}
  E.~Witten,
  ``Anti-de Sitter space and holography,''
  Adv.\ Theor.\ Math.\ Phys.\  {\bf 2}, 253 (1998)
  [hep-th/9802150].





\bibitem{Babington:2003vm}
  J.~Babington, J.~Erdmenger, N.~J.~Evans, Z.~Guralnik and I.~Kirsch,
  ``Chiral symmetry breaking and pions in nonsupersymmetric gauge / gravity duals,''
  Phys.\ Rev.\ D {\bf 69}, 066007 (2004)
  [hep-th/0306018].

\bibitem{Kruczenski:2003be}
  M.~Kruczenski, D.~Mateos, R.~C.~Myers and D.~J.~Winters,
  ``Meson spectroscopy in AdS / CFT with flavor,''
  JHEP {\bf 0307}, 049 (2003)
  [hep-th/0304032].

\bibitem{Kruczenski:2003uq}
  M.~Kruczenski, D.~Mateos, R.~C.~Myers and D.~J.~Winters,
  ``Towards a holographic dual of large N(c) QCD,''
  JHEP {\bf 0405}, 041 (2004)
  [hep-th/0311270].

\bibitem{Sakai:2004cn}
  T.~Sakai and S.~Sugimoto,
  ``Low energy hadron physics in holographic QCD,''
  Prog.\ Theor.\ Phys.\  {\bf 113}, 843 (2005)
  [hep-th/0412141].

\bibitem{Sakai:2005yt}
  T.~Sakai and S.~Sugimoto,
  ``More on a holographic dual of QCD,''
  Prog.\ Theor.\ Phys.\  {\bf 114}, 1083 (2005)
  [hep-th/0507073].

\bibitem{Huang:2007fv}
  S.~He, M.~Huang, Q.~S.~Yan and Y.~Yang,
  ``Confront Holographic QCD with Regge Trajectories,''
  Eur.\ Phys.\ J.\ C {\bf 66}, 187 (2010)
  [arXiv:0710.0988 [hep-ph]].

\bibitem{Abt:2019tas}
R.~Abt, J.~Erdmenger, N.~Evans and K.~S.~Rigatos,
JHEP \textbf{11} (2019), 160
[arXiv:1907.09489 [hep-th]].

\bibitem{Nakas:2020hyo}
T.~Nakas and K.~S.~Rigatos,
``Fermions and baryons as open-string states from brane junctions,''
[arXiv:2010.00025 [hep-th]].

\bibitem{Gubser:2008ny}
  S.~S.~Gubser and A.~Nellore,
  ``Mimicking the QCD equation of state with a dual black hole,''
  Phys.\ Rev.\ D {\bf 78} (2008) 086007
  [arXiv:0804.0434 [hep-th]].

\bibitem{Gubser:2008yx}
  S.~S.~Gubser, A.~Nellore, S.~S.~Pufu and F.~D.~Rocha,
  ``Thermodynamics and bulk viscosity of approximate black hole duals to finite temperature quantum chromodynamics,''
  Phys.\ Rev.\ Lett.\  {\bf 101} (2008) 131601
  [arXiv:0804.1950 [hep-th]].

\bibitem{DeWolfe:2010he}
  O.~DeWolfe, S.~S.~Gubser and C.~Rosen,
  ``A holographic critical point,''
  Phys.\ Rev.\ D {\bf 83} (2011) 086005
  [arXiv:1012.1864 [hep-th]].

\bibitem{Gursoy:2007cb}
  U.~Gursoy and E.~Kiritsis,
  ``Exploring improved holographic theories for QCD: Part I,''
  JHEP {\bf 0802} (2008) 032
  [arXiv:0707.1324 [hep-th]].

\bibitem{Gursoy:2007er}
  U.~Gursoy, E.~Kiritsis and F.~Nitti,
  ``Exploring improved holographic theories for QCD: Part II,''
  JHEP {\bf 0802} (2008) 019
  [arXiv:0707.1349 [hep-th]].

\bibitem{Finazzo:2014cna}
  S.~I.~Finazzo, R.~Rougemont, H.~Marrochio and J.~Noronha,
  ``Hydrodynamic transport coefficients for the non-conformal quark-gluon plasma from holography,''
  JHEP {\bf 1502} (2015) 051
  [arXiv:1412.2968 [hep-ph]].

\bibitem{Zollner:2018uep}
  R.~Zollner and B.~Kampfer,
  ``Phase structures emerging from holography with Einstein gravity -- dilaton models at finite temperature,''
  Eur.\ Phys.\ J.\ Plus {\bf 135} (2020) no.3,  304
  [arXiv:1807.04260 [hep-th]].




\bibitem{Ballon-Bayona:2020xls}
  A.~Ballon-Bayona, H.~Boschi-Filho, E.~Folco Capossoli and D.~M.~Rodrigues,
  ``Criticality from EMD holography at finite temperature and density,''
  arXiv:2006.08810 [hep-th].

\bibitem{Bohra:2020qom}
  H.~Bohra, D.~Dudal, A.~Hajilou and S.~Mahapatra,
  ``Chiral transition in the probe approximation from an Einstein-Maxwell-dilaton gravity model,''
  arXiv:2010.04578 [hep-th].

\bibitem{Mamani:2020pks}
  L.~A.~H.~Mamani, C.~V.~Flores and V.~T.~Zanchin,
  ``Phase diagram and compact stars in a holographic QCD model,''
  Phys.\ Rev.\ D {\bf 102} (2020) no.6,  066006
  [arXiv:2006.09401 [hep-th]].

\bibitem{He:2020fdi}
  S.~He, Y.~Yang and P.~H.~Yuan,
  ``Analytic Study of Magnetic Catalysis in Holographic QCD,''
  arXiv:2004.01965 [hep-th].

\bibitem{Colangelo:2011sr}
  P.~Colangelo, F.~Giannuzzi, S.~Nicotri and V.~Tangorra,
  ``Temperature and quark density effects on the chiral condensate: An AdS/QCD study,''
  Eur.\ Phys.\ J.\ C {\bf 72} (2012) 2096  [arXiv:1112.4402 [hep-ph]].  

\bibitem{Dudal:2015wfn}
  D.~Dudal, D.~R.~Granado and T.~G.~Mertens,
  ``No inverse magnetic catalysis in the QCD hard and soft wall models,''
  Phys.\ Rev.\ D {\bf 93}, no. 12, 125004 (2016)
  [arXiv:1511.04042 [hep-th]].

\bibitem{Chelabi:2015cwn}
  K.~Chelabi, Z.~Fang, M.~Huang, D.~Li and Y.~L.~Wu,
  ``Realization of chiral symmetry breaking and restoration in holographic QCD,''
  Phys.\ Rev.\ D {\bf 93}, no. 10, 101901 (2016) [arXiv:1511.02721 [hep-ph]].

\bibitem{Chelabi:2015gpc}
  K.~Chelabi, Z.~Fang, M.~Huang, D.~Li and Y.~L.~Wu,
  ``Chiral Phase Transition in the Soft-Wall Model of AdS/QCD,''
  JHEP {\bf 1604} (2016) 036
  [arXiv:1512.06493 [hep-ph]].

\bibitem{Fang:2015ytf}
  Z.~Fang, S.~He and D.~Li,
  ``Chiral and Deconfining Phase Transitions from Holographic QCD Study,''
  Nucl.\ Phys.\ B {\bf 907} (2016) 187
  [arXiv:1512.04062 [hep-ph]].

\bibitem{Li:2016gfn}
  D.~Li, M.~Huang, Y.~Yang and P.~H.~Yuan,
  ``Inverse Magnetic Catalysis in the Soft-Wall Model of AdS/QCD,''
  JHEP {\bf 1702} (2017) 030
  [arXiv:1610.04618 [hep-th]].

\bibitem{Li:2016smq}
  D.~Li and M.~Huang,
  ``Chiral phase transition of QCD with $N_f=2+1$ flavors from holography,''
  JHEP {\bf 1702} (2017) 042
  [arXiv:1610.09814 [hep-ph]].

\bibitem{Bartz:2016ufc}
  S.~P.~Bartz and T.~Jacobson,
  ``Chiral Phase Transition and Meson Melting from AdS/QCD,''
  Phys.\ Rev.\ D {\bf 94} (2016) 075022
  [arXiv:1607.05751 [hep-ph]].

\bibitem{Fang:2016nfj}
  Z.~Fang, Y.~L.~Wu and L.~Zhang,
  ``Chiral phase transition and meson spectrum in improved soft-wall AdS/QCD,''
  Phys.\ Lett.\ B {\bf 762} (2016) 86
  [arXiv:1604.02571 [hep-ph]].

\bibitem{Bartz:2017jku}
  S.~P.~Bartz and T.~Jacobson,
  ``Chiral phase transition at finite chemical potential in 2+1 -flavor soft-wall anti–de Sitter space QCD,''
  Phys.\ Rev.\ C {\bf 97} (2018) no.4,  044908
  [arXiv:1801.00358 [hep-ph]].

\bibitem{Fang:2018vkp}
  Z.~Fang, Y.~L.~Wu and L.~Zhang,
  ``Chiral Phase Transition with 2+1 quark flavors in an improved soft-wall AdS/QCD Model,''
  arXiv:1805.05019 [hep-ph].

\bibitem{Gherghetta-Kapusta-Kelley}
T.~Gherghetta, J.~I.~Kapusta and T.~M.~Kelley,
  ``Chiral symmetry breaking in the soft-wall AdS/QCD model,''
Phys.\ Rev.\ D {\bf 79} (2009) 076003;
 [arXiv:0902.1998 [hep-ph]].  

\bibitem{Gherghetta-Kapusta-Kelley-2}
T.~M.~Kelley, S.~P.~Bartz and J.~I.~Kapusta,
  ``Pseudoscalar Mass Spectrum in a Soft-Wall Model of AdS/QCD,''
  Phys.\ Rev.\ D {\bf 83} (2011) 016002;
[arXiv:1009.3009 [hep-ph]].  

\bibitem{Li:2012ay}
  D.~Li, M.~Huang and Q.~S.~Yan,
  ``A dynamical soft-wall holographic QCD model for chiral symmetry breaking and linear confinement,''
  Eur.\ Phys.\ J.\ C {\bf 73} (2013) 2615
  [arXiv:1206.2824 [hep-th]].




\bibitem{Li:2013oda}
  D.~Li and M.~Huang,
  ``Dynamical holographic QCD model for glueball and light meson spectra,''
  JHEP {\bf 1311} (2013) 088
  [arXiv:1303.6929 [hep-ph]].



\bibitem{YLWu}
  Y.~-Q.~Sui, Y.~-L.~Wu, Z.~-F.~Xie and Y.~-B.~Yang,
  ``Prediction for the Mass Spectra of Resonance Mesons in the Soft-Wall AdS/QCD with a Modified 5D Metric,''
  Phys.\ Rev.\ D {\bf 81} (2010) 014024;
   [arXiv:0909.3887 [hep-ph]].  

\bibitem{YLWu-1}
Y.~-Q.~Sui, Y.~-L.~Wu and Y.~-B.~Yang,
  ``Predictive AdS/QCD Model for Mass Spectra of Mesons with Three Flavors,''
  Phys.\ Rev.\ D {\bf 83} (2011) 065030.
  [arXiv:1012.3518 [hep-ph]].  





\bibitem{Colangelo:2008us}
  P.~Colangelo, F.~De Fazio, F.~Giannuzzi, F.~Jugeau and S.~Nicotri,
  ``Light scalar mesons in the soft-wall model of AdS/QCD,''
   Phys.\ Rev.\ D {\bf 78}, 055009 (2008)  [arXiv:0807.1054 [hep-ph]].  

\bibitem{Ballon-Bayona:2020qpq}
  A.~Ballon-Bayona and L.~A.~H.~Mamani,
  ``Nonlinear realization of chiral symmetry breaking in holographic soft wall models,''
  Phys.\ Rev.\ D {\bf 102} (2020) no.2,  026013
  [arXiv:2002.00075 [hep-ph]].

\bibitem{FolcoCapossoli:2019imm}
  E.~Folco Capossoli, M.~A.~Martín Contreras, D.~Li, A.~Vega and H.~Boschi-Filho,
  ``Hadronic Spectra from Deformed AdS Backgrounds,''
  Chin.\ Phys.\ C {\bf 44} (2020) no.6,  064104
  [arXiv:1903.06269 [hep-ph]].

\bibitem{Contreras:2018hbi}
M.~\'A.~Mart\'\i{}n Contreras, A.~Vega and S.~Cort\'es,
Chin. J. Phys. \textbf{66} (2020), 715-723
[arXiv:1811.10731 [hep-ph]].

\bibitem{Karch:2006pv}
  A.~Karch, E.~Katz, D.~T.~Son and M.~A.~Stephanov,
  ``Linear confinement and AdS/QCD,''
  Phys.\ Rev.\ D {\bf 74}, 015005 (2006)
  [hep-ph/0602229].


\bibitem{Chen:2019rez}
  X.~Chen, D.~Li, D.~Hou and M.~Huang,
  ``Quarkyonic phase from quenched dynamical holographic QCD model,''
  JHEP {\bf 2003}, 073 (2020)
  [arXiv:1908.02000 [hep-ph]].

\bibitem{Yang:2014bqa}
  Y.~Yang and P.~H.~Yuan,
  ``A Refined Holographic QCD Model and QCD Phase Structure,''
  JHEP {\bf 1411}, 149 (2014)
  [arXiv:1406.1865 [hep-th]].


\bibitem{Dudal:2017max}
D.~Dudal and S.~Mahapatra,
``Thermal entropy of a quark-antiquark pair above and below deconfinement from a dynamical holographic QCD model,''
Phys. Rev. D \textbf{96} (2017) no.12, 126010
[arXiv:1708.06995 [hep-th]].


\bibitem{Cai:2012xh}
  R.~G.~Cai, S.~He and D.~Li,
  ``A hQCD model and its phase diagram in Einstein-Maxwell-Dilaton system,''
  JHEP {\bf 1203}, 033 (2012)
  [arXiv:1201.0820 [hep-th]].

\bibitem{Yang:2015aia}
Y.~Yang and P.~H.~Yuan,
``Confinement-deconfinement phase transition for heavy quarks in a soft wall holographic QCD model,''
JHEP \textbf{12} (2015), 161
[arXiv:1506.05930 [hep-th]].

\bibitem{deForcrand:2002hgr}
P.~de Forcrand and O.~Philipsen,
``The QCD phase diagram for small densities from imaginary chemical potential,''
Nucl. Phys. B \textbf{642} (2002), 290-306
[arXiv:hep-lat/0205016 [hep-lat]].

\bibitem{McInnes:2014haa}
  B.~McInnes,
  ``Angular Momentum in QGP Holography,''
  Nucl.\ Phys.\ B {\bf 887}, 246 (2014)
  [arXiv:1403.3258 [hep-th]].



\bibitem{McInnes:2018pmk}
  B.~Mcinnes,
  ``Holography of Low-Centrality Heavy Ion Collisions,''
  Int.\ J.\ Mod.\ Phys.\ A {\bf 34} (2019) no.29,  1950174
  [arXiv:1805.09558 [hep-th]].

\bibitem{McInnes:2018mwj}
  B.~McInnes,
  ``Viscosity vs. Vorticity in the Quark-Gluon Plasma,''
  Nucl.\ Phys.\ B {\bf 953} (2020) 114951
  [arXiv:1812.07146 [hep-th]].

\bibitem{Arefeva:2020jvo}
  I.~Y.~Aref'eva, A.~A.~Golubtsova and E.~Gourgoulhon,
  ``Holographic drag force in 5d Kerr-AdS black hole,''
  arXiv:2004.12984 [hep-th].


\bibitem{Erices:2017izj}
C.~Erices and C.~Martinez,
``Rotating hairy black holes in arbitrary dimensions,''
Phys. Rev. D \textbf{97}, no.2, 024034 (2018)
[arXiv:1707.03483 [hep-th]].


\bibitem{BravoGaete:2017dso}
  M.~Bravo Gaete, L.~Guajardo and M.~Hassaine,
  ``A Cardy-like formula for rotating black holes with planar horizon,''
  JHEP {\bf 1704}, 092 (2017)
  [arXiv:1702.02416 [hep-th]].

\bibitem{Sheykhi:2010pya}
A.~Sheykhi and S.~Hendi,
``Charged rotating dilaton black branes in AdS universe,''
Gen. Rel. Grav. \textbf{42}, 1571-1583 (2010)
[arXiv:0911.2831 [hep-th]].


\bibitem{Awad:2002cz}
  A.~M.~Awad,
  ``Higher dimensional charged rotating solutions in (A)dS space-times,''
  Class.\ Quant.\ Grav.\  {\bf 20}, 2827 (2003)
  [hep-th/0209238].


\bibitem{Nadi:2019bqu}
H.~Nadi, B.~Mirza, Z.~Sherkatghanad and Z.~Mirzaiyan,
``Holographic entanglement first law for $d$ + 1 dimensional rotating cylindrical black holes,''
Nucl. Phys. B \textbf{949} (2019), 114822
[arXiv:1904.11344 [gr-qc]].

\bibitem{Li:2011hp}
  D.~Li, S.~He, M.~Huang and Q.~S.~Yan,
  ``Thermodynamics of deformed AdS$_5$ model with a positive/negative quadratic correction in graviton-dilaton system,''
  JHEP {\bf 1109}, 041 (2011)
  [arXiv:1103.5389 [hep-th]].


\bibitem{Kajantie:2011nx}
K.~Kajantie, M.~Krssak, M.~Vepsalainen and A.~Vuorinen,
``Frequency and wave number dependence of the shear correlator in strongly coupled hot Yang-Mills theory,''
Phys. Rev. D \textbf{84} (2011), 086004
doi:10.1103/PhysRevD.84.086004
[arXiv:1104.5352 [hep-ph]].


\bibitem{Farakos:2009fx}
K.~Farakos, A.~P.~Kouretsis and P.~Pasipoularides,
``Anti de Sitter 5D black hole solutions with a self-interacting bulk scalar field: A Potential reconstruction approach,''
Phys. Rev. D \textbf{80} (2009), 064020
[arXiv:0905.1345 [hep-th]].


\bibitem{Cai:2012eh}
R.~G.~Cai, S.~Chakrabortty, S.~He and L.~Li,
``Some aspects of QGP phase in a hQCD model,''
JHEP \textbf{02} (2013), 068
[arXiv:1209.4512 [hep-th]].



\bibitem{Zheng1983Hawking}
   Z.~Zhao and Y.~G,
``Hawking radiation in a four-dimensional pseudo-riemannian stationary spacetime,''
Chinese Astronomy \& Astrophysics(1983)

\bibitem{Natsuume:2014sfa}
M.~Natsuume,
``AdS/CFT Duality User Guide,''
Lect. Notes Phys. \textbf{903} (2015), pp.39
[arXiv:1409.3575 [hep-th]].

\bibitem{Cvetic:1999ne}
M.~Cvetic and S.~S.~Gubser,
``Phases of R charged black holes, spinning branes and strongly coupled gauge theories,''
JHEP \textbf{04} (1999), 024
[arXiv:hep-th/9902195 [hep-th]].

\bibitem{Caldarelli:1999xj}
M.~M.~Caldarelli, G.~Cognola and D.~Klemm,
``Thermodynamics of Kerr-Newman-AdS black holes and conformal field theories,''
Class. Quant. Grav. \textbf{17} (2000), 399-420
[arXiv:hep-th/9908022 [hep-th]].



\bibitem{Burger:2014xga}
F.~Burger \textit{et al.} [tmfT],
``Equation of state of quark-gluon matter from lattice QCD with two flavors of twisted mass Wilson fermions,''
Phys. Rev. D \textbf{91} (2015) no.7, 074504
[arXiv:1412.6748 [hep-lat]].





\bibitem{Boyd:1996bx}
G.~Boyd, J.~Engels, F.~Karsch, E.~Laermann, C.~Legeland, M.~Lutgemeier and B.~Petersson,
``Thermodynamics of SU(3) lattice gauge theory,''
Nucl. Phys. B \textbf{469}, 419-444 (1996)
[arXiv:hep-lat/9602007 [hep-lat]].

\bibitem{Dehghani:2002rr}
M.~Dehghani,
``Thermodynamics of rotating charged black strings and (A)dS / CFT correspondence,''
Phys. Rev. D \textbf{66}, 044006 (2002)
[arXiv:hep-th/0205129 [hep-th]].


\bibitem{He:2013qq}
S.~He, S.~Wu, Y.~Yang and P.~Yuan,
``Phase Structure in a Dynamical Soft-Wall Holographic QCD Model,''
JHEP \textbf{04} (2013), 093
[arXiv:1301.0385 [hep-th]].

\bibitem{Abuki:2008nm}
H.~Abuki, R.~Anglani, R.~Gatto, G.~Nardulli and M.~Ruggieri,
``Chiral crossover, deconfinement and quarkyonic matter within a Nambu-Jona Lasinio model with the Polyakov loop,''
Phys. Rev. D \textbf{78}, 034034 (2008)
[arXiv:0805.1509 [hep-ph]].


\bibitem{Giataganas:2012zy}
D.~Giataganas,
``Probing strongly coupled anisotropic plasma,''
JHEP \textbf{07} (2012), 031
[arXiv:1202.4436 [hep-th]].

\bibitem{Giataganas:2017koz}
D.~Giataganas, U.~G\"ursoy and J.~F.~Pedraza,
``Strongly-coupled anisotropic gauge theories and holography,''
Phys. Rev. Lett. \textbf{121} (2018) no.12, 121601
[arXiv:1708.05691 [hep-th]].






\bibitem{Andreev:2006nw}
  O.~Andreev and V.~I.~Zakharov,
  ``On Heavy-Quark Free Energies, Entropies, Polyakov Loop, and AdS/QCD,''
  JHEP {\bf 0704}, 100 (2007)
  [hep-ph/0611304].


\bibitem{Maldacena:1998im}
J.~M.~Maldacena,
``Wilson loops in large N field theories,''
Phys. Rev. Lett. \textbf{80} (1998), 4859-4862
[arXiv:hep-th/9803002 [hep-th]].


\bibitem{Rey:1998ik}
S.~Rey and J.~Yee,
``Macroscopic strings as heavy quarks in large N gauge theory and anti-de Sitter supergravity,''
Eur. Phys. J. C \textbf{22} (2001), 379-394
[arXiv:hep-th/9803001 [hep-th]].

\bibitem{Ewerz:2016zsx}
C.~Ewerz, O.~Kaczmarek and A.~Samberg,
``Free Energy of a Heavy Quark-Antiquark Pair in a Thermal Medium from AdS/CFT,''
JHEP \textbf{03} (2018), 088
[arXiv:1605.07181 [hep-th]].





\bibitem{Colangelo:2010pe}
  P.~Colangelo, F.~Giannuzzi and S.~Nicotri,
  ``Holography, Heavy-Quark Free Energy, and the QCD Phase Diagram,''
  Phys.\ Rev.\ D {\bf 83}, 035015 (2011)
  [arXiv:1008.3116 [hep-ph]].


\bibitem{Chen:2017lsf}
  X.~Chen, S.~Q.~Feng, Y.~F.~Shi and Y.~Zhong,
  ``Moving heavy quarkonium entropy, effective string tension, and the QCD phase diagram,''
  Phys.\ Rev.\ D {\bf 97}, no. 6, 066015 (2018)
  [arXiv:1710.00465 [hep-ph]].

\bibitem{Polyakov:1978vu}
A.~M.~Polyakov,
``Thermal Properties of Gauge Fields and Quark Liberation,''
Phys. Lett. B \textbf{72} (1978), 477-480

\bibitem{Bali:1993tz}
G.~Bali, J.~Fingberg, U.~M.~Heller, F.~Karsch and K.~Schilling,
``The Spatial string tension in the deconfined phase of the (3+1)-dimensional SU(2) gauge theory,''
Phys. Rev. Lett. \textbf{71} (1993), 3059-3062
[arXiv:hep-lat/9306024 [hep-lat]].




\end{thebibliography}
\end{document}